\documentclass[reprint,amsmath,amssymb,aps,prb,twocolumn,float]{revtex4-2}
\usepackage{graphicx}
\usepackage{graphics}
\usepackage{dcolumn}
\usepackage{bm}
\usepackage{amsfonts}
\usepackage{amssymb}
\usepackage{amsmath}
\usepackage{physics}
\usepackage{float}
\usepackage{xcolor}
\usepackage{soul}
\usepackage{natbib}
\usepackage{braket}
\usepackage{hyperref}
\usepackage{breqn}
\usepackage{makecell}
\usepackage{feynmp-auto}
\usepackage{subfigure}

\graphicspath{ {Feynman_diagrams/} }
\newcommand{\vk}{{\mathbf{k}}}

\definecolor{green}{rgb}{0,0.6,0.1}

\newcommand{\NCTS}{Physics Division, National Center for Theoretical Sciences, Taipei 10617, Taiwan\looseness=-1}
\newcommand{\IAMS}{Institute of Atomic and Molecular Sciences, Academia Sinica, Taipei 10617, Taiwan\looseness=-1}

\begin{document}
\title{Diagrammatic approach to excitonic effects on nonlinear optical response}

\author{Yu-Tzu Chang$^{1}$}
\author{Yang-Hao Chan$^{1,2}$}
\email{yanghao@gate.sinica.edu.tw}
\affiliation{$^{1}$\IAMS}
\affiliation{$^{2}$\NCTS}
\date{\today}
\begin{abstract}
    Optical responses of atomically thin 2D materials are greatly influenced by electron-hole interactions. It is by far established that exciton signatures can be well-identified in the optical absorption spectrum of quasi-2D materials. However, the same level of understanding of excitonic effects on nonlinear optical responses and the ability to compute them accurately is still much desired.
    Based on the functional integral formalisms and working in the velocity gauge, we introduce a convenient Feynman diagram approach for calculating nonlinear responses including excitonic effects. 
    By dressing electron-photon interactions with electron-hole ladder diagrams, we derive an expression for second-order optical responses and provide a comprehensive description of excitonic effects. We apply our approach to a monolayer h-BN model and show qualitative changes in the second harmonic generation spectrum when comparing with results assuming independent particles. Our approach can be readily extended to higher order optical responses and is feasible for first-principles calculations. 
\end{abstract}

\maketitle

\section{introduction}
Modern experimental techniques integrated with theoretical simulations enable precise measurements and control of optical responses, facilitating the validation of quantum material models. 
To accurately describe optical responses in solid-state systems, significant progress has been made in the past few decades. Theoretically, light-matter interactions can be written in terms of the vector potential and electron momentum in a minimal-coupling scheme, which is often called the velocity gauge. Alternatively, the coupling can also be cast as the product of the electron position operator and the electric field, which is known as the length gauge. A well-defined light-matter interaction in periodic systems was established by Blount~\cite{blount}, which later led to the development of modern polarization theory in terms of Berry connection~\cite{King-Smith1993,Resta1994}. Based on Blount's treatment, Sipe et al., derived nonlinear optical responses in terms of density matrix formalism through perturbation expansion of light-matter couplings~\cite{Sipe2000}. A systematic comparison of these two gauges is made and it has been shown that the two are formally equivalent~\cite{Ventura2017,Taghizadeh2017}. Applications of both treatments to study optical responses of real materials within the independent particle approximation (IPA) have been widely conducted.

Nonlinear optical responses describe a plethora of phenomena in which the light-induced polarization or current does not scale linearly with the electric field strength of the incident light. Typical examples of nonlinear responses include harmonic generation, optical rectification, shift current and frequency mixing effects, etc.,~\cite{Boyd2008} many of which have practical applications. Recent investigations on nonlinear optical spin Hall conductivity and nonlinear anomalous Hall effect~\cite{Taghizadeh_2020,Li2021} further reveal the potential applications in spintronics devices.
Aside from applications in optoelectronic devices, studies of nonlinear optical response have also advanced our fundamental understanding of light-matter interactions.
Recent theoretical study on nonlinear optical responses culminates with the establishment of its connection to quantum geometry~\cite{L.Fu2023arXiv, Yan2020, Ahn2022} and topology~\cite{Nagaosa2016}.
Notably, the role of electron-hole correlations hence excitonic effects is largely ignored owing to the typical small exciton binding energy in 3D bulk semiconductors.

The ability to prepare atomically thin 2D materials has brought opportunities to study excitonic effects on optical responses as it is well-known by now that excitonic effects become particularly strong due to reduced screening and quantum confinement effects in quasi-2D materials. 
Combining efforts of experimental and theoretical first-principles investigations have shown that bound excitons with binding energies of hundreds of meV can be clearly identified in optical absorption or photoluminescence spectrum~\cite{Qiu2013, Chernikov2014,Wang2018}. However, our understanding of excitonic effects on nonlinear optical responses is far from complete as the role of excitons on the nonlinear optical spectrum is still under debate~\cite{Malard2013,Wang2015,Lafeta_2021} and accurate $\textit{ab initio}$ computational tools are still under development~\cite{Chang2001,Leitsmann2005, luppi_review_2016,Attaccalite2013,Gruning2014,Chan2021,Huang2023,Quek2023}.

Excitonic effects on optical responses have been studied by treating electron-hole interactions on the mean-field type of approximation with either Hartree-Fock, screened exchange self-energy or range-separated hybrid exchange-correlation potentials~\cite{luppi_review_2016,Attaccalite2013,Gruning2014,Chan2021}. In the wave function-based method, real-time propagation approaches have been conducted to study correlation effects on nonlinear optical responses. In a recent work, excitonic effects were considered in the framework of dynamical Berry phase polarization where the light-matter coupling is written in the length gauge. Expressions of general second order optical responses including excitonic effects in both length and velocity gauge \cite{Quek2023} were derived with the density matrix formalism. In particular, velocity and position operators are defined in the exciton basis to ensure the gauge invariance~\cite{Taghizadeh2018}.

Although the perturbative derivation within the density matrix formalism is conceptually clean, the derivation involves tedious book-keeping of various orders of perturbations even in the IPA. To provide a concise physical interpretation of high order optical responses, a recent work demonstrates a Feynman diagram approach to calculate nonlinear optical responses in the velocity gauge within IPA~\cite{Parker2019}.
An extension of this method to spatially dispersive nonlinear responses was reported~\cite{Mele2023arXiv}.
This motivates us to pursue a diagrammatic approach for the derivation of nonlinear optical responses with excitonic effects.

In the practical calculation, excitons are solutions of the Bethe-Salpeter equation (BSE) within a static approximation to screened Coulomb interactions.
Diagrammatically, BSE can be derived from the equation of motion of electron-hole correlations, in which a series of interactions between electrons and holes is written as the so-called ladder diagrams.
In Ref.~\cite{Sven2020}, incorporations of ladder diagrams into Raman scattering intensity were considered to derive the excitonic effects on resonant Raman spectroscopy of layered materials.
The derived expression can be naturally written in terms of exciton-phonon coupling matrix elements introduced in other contexts~\cite{Antonius2022,Chen2020}.
However, ladder diagrams' incorporation into nonlinear optical responses has not yet been reported.\par
In this work, we develop a convenient Feynman diagram approach for nonlinear optical responses including excitonic effects. Our goal is to derive a general expression for second-order optical response with excitonic effects using the diagrammatic approach in the velocity gauge. The resulting expression offers practical advantages, revealing the physical interpretations of seemingly complicated summations over matrix elements and readily distinguishing one-, two-, and three-photon processes. Moreover, the expression can be straightforwardly implemented for tight-binding models and for first-principles calculations of real materials.
The rest of the paper is structured as follows: In Sec.~\ref{sec:setup}, we revisit the functional derivative formalism and Feynman rules for the derivation of the linear and second-order optical conductivity in the IPA. We introduce the electron-hole correlation function, ladder diagrams, and BSE for electron-photon coupling vertex in Sec.~\ref{sec:eh_diagrams}. By employing the vertex correction method, we derive the expression of the first order and the second order optical responses with excitonic effects in Sec.~\ref{sec:optical_with_eh}. In Sec.~\ref{sec:model}, we apply our approach to an effective two-band tight-binding model for a monolayer h-BN, along with a comparison with other derivations. We conclude and discuss the outlook of our approach in Sec.~\ref{sec:conclusion}.

\section{Functional integral setup}
\label{sec:setup}
A detailed introduction of the functional integral approach can be found in Ref.~\cite{Peskin:1995ev,Parker2019} and references therein. In this section, we revisit the key equations and introduce our notation to set the stage for the derivations in the next section.

To determine the optical conductivity using the functional integral formalism, we start by writing the partition function in the form of a path integral.
\begin{equation}
    Z =\int D[c,\bar{c}] e^{-i\int dt H(t)}, \nonumber
\label{eq:partition func}
\end{equation}
where $c$, $\bar{c}$ are Grassman fields. The Hamiltonian of the system subjected to a time-dependent optical field is represented in the second quantization basis as
\begin{equation}
    \hat{H}(t) = \hat{H}_{0}+\hat{H}_{A}(t), \nonumber
\label{eq:H(t)}
\end{equation}
where $\hat{H}_{0}$ denotes the unperturbed and independent particle part of the Hamiltonian and $\hat{H}_A(t)$ describes the light-matter coupling. 
We have
\begin{equation}
    \hat{H}_{0} = \int d\vk\  \varepsilon_{ab\vk} c^{\dagger}_{\vk a} c_{\vk b}, \nonumber
\label{eq:H_0}
\end{equation}
where we define the electron creation and annihilation operators $c^{\dagger}_{\vk a}$ and $c_{\vk b}$ with crystal momentum $\vk$ and band indices $a$ and $b$, respectively; $\varepsilon_{ab}$ is the matrix element of the unperturbed Hamiltonian.

The interaction part of the Hamiltonian, $\hat{H}_{A}(t)$ is written in the velocity gauge.
Following Ref.~\cite{235446, Parker2019}, the light-matter couplings can be written as an expansion of the unperturbed $H_0$ in powers of the vector potential field $A^{\alpha}$,
\begin{equation}
\begin{split}
    \hat{H}_{A}(t)& =e A^{\alpha_{1}}(t) \hat{h}^{\alpha_{1}} + \frac{e^{2}}{2}A^{\alpha_{1}}(t)A^{\alpha_{2}}(t) \hat{h}^{\alpha_{1}\alpha_{2}}+ \cdots
    \\& = \sum_{n=1}^{\infty} \frac{e^{n}}{n!}A^{\alpha_{1}}(t)A^{\alpha_{2}}(t)\cdots A^{\alpha_{n}}(t)\hat{h}^{\alpha_{1}\alpha_{2}\cdots \alpha_{n}}, \nonumber
\end{split}
\label{eq:H_{A}}
\end{equation}
where $e$ is the electron charge and $\alpha$ is the Cartesian index of the polarization direction of the external field. The $n$-th order derivative of $\hat{h}$ is defined as 
\begin{equation}
\hat{h}^{\alpha_{1} \alpha_{2}\cdots \alpha_{n}}=\hbar^{-n}D^{\alpha_{n}}\cdots D^{\alpha_{2}}D^{\alpha_{1}}\hat{H}_{0}, \nonumber
\label{eq:h_hat}
\end{equation}
where $D^{\alpha}$ is the covariant derivative operator and its operation is defined through the commutator with another operator, $[D^{\alpha},\hat{O}]_{ab} = \frac{\partial O_{ab}}{\partial k^{\alpha}}-i[\xi^{\alpha}, \hat{O}]_{ab}$.
At the lowest order, the velocity operator in the eigen-basis of the unperturbed Hamiltonian reads,
\begin{equation}
    h^{\alpha}_{ab}=\frac{1}{\hbar}[D^{\alpha},\hat{H}_{0}]_{ab} = \frac{1}{\hbar}\frac{\partial \varepsilon_{ab\vk}}{\partial k^{\alpha}}\delta_{ab}-i\frac{\xi_{ab}^{\alpha}}{\hbar}(\varepsilon_{a}-\varepsilon_{b}),
\label{eq:h_ab}
\end{equation}
where $\xi_{ab}^{\alpha}$ is the matrix element of the interband Berry connection and we hide the crystal momentum $\vk$ to make the notation clean when there is no ambiguity. The lowest order light-matter coupling term $H_{A}(t)$ is the familiar form, $e\int \hat{\boldsymbol{v}}_0 \cdot \boldsymbol{A}(t)$.

To work with the electric field directly, we employ $E^\alpha(\omega) = i\omega A^\alpha(\omega)$ to convert $\hat{H}_{A}$ into $\hat{H}_{E}$ using the Fourier transformation,
\begin{equation}
    \hat{H}_{E}(t) = \sum_{n=1}^{\infty} \frac{e^{n}}{n!} \prod_{l=1}^{n}\int d\omega_{l}e^{-i\omega_{l}t} \frac{E^{\alpha_{l}}(\omega_{l})}{i\omega_{l}}\hat{h}^{\alpha_{1}\cdots \alpha_{n}}.
    \label{eq:H_E}
\end{equation}
The current density is calculated as the expectation value of the current density operator $\hat{\boldsymbol{J}}$. In terms of the partition function, we have 
\begin{equation}
    \expval{\hat{J}^{\mu}(t)} = \frac{1}{Z}\int D[c,\Bar{c}]e\hat{v}^{\mu}(t) e^{-i\int dt'(H_{0}+H_{E}(t'))},
\label{eq:J}
\end{equation}
where $\hat{v}^{\mu}(t)=D^{\mu}[\hat{H}_{0}+\hat{H}_{E}(t)]$, denotes the time-dependent velocity operator obtained by taking the derivative of the total Hamiltonian. Explicitly, it reads,
\begin{equation}
    \hat{v}^{\mu}(t) = \sum_{n=0}^{\infty} \frac{e^{n}}{n!} \prod_{l=1}^{n}\int d\omega_{l}e^{-i\omega_{l}t} \frac{E^{\alpha_{l}}(\omega_{l})}{i\omega_{l}}\hat{h}^{\mu \alpha_{1}\cdots \alpha_{n}}
    \label{eq:v_t}
\end{equation}
Therefore, both $\hat{\boldsymbol{v}}(t)$ and $\hat{H}_{E}(t)$ are functionals of the electric field $\boldsymbol{E}(t)$.
In the frequency domain, we define conductivity tensors which are related to the current density as~\cite{235446}
\begin{align}
    J^{\mu}(\omega) & =\int\frac{d\omega_{1}}{2\pi}\sigma^{\mu\alpha}(\omega_{1})E^{\alpha}(\omega_{1})(2\pi)\delta(\omega-\omega_{1})\nonumber\\
 & +\int\frac{d\omega_{1}}{2\pi}\frac{d\omega_{2}}{2\pi}\sigma^{\mu\alpha\beta}(\omega_{1},\omega_{2})E^{\alpha}(\omega_{1})E^{\beta}(\omega_{2})\nonumber\\
 &\times(2\pi)\delta(\omega-\omega_{1}-\omega_{2})+...
 \label{eq:defJw}
\end{align}

As a demonstration of the functional derivative approach and an introduction of notations and diagrams, we reproduce the derivation of the linear optical conductivity tensor given in Ref.~\cite{Parker2019} below. The derivation of the second order conductivity tensor will be given in Appendix~\ref{sec:appendixB1}.

At the linear order, we compute the conductivity tensor by taking the functional derivative of $J(t)$ with respect to $E(t)$ and performing a Fourier transformation, 
\begin{align}
    &\sigma^{\mu\alpha}(\omega,\omega_1)\delta(\omega-\omega_1)=\frac{\delta J^\mu(\omega)}{\delta E^\alpha(\omega_1)} \nonumber\\
    & =\int \frac{dt_1}{2\pi} e^{-i\omega_1t_1} \frac{\delta}{\delta E^\alpha(t_1)}\int e^{i\omega t}dt J^\mu(t) \nonumber \\
    & = \int \frac{dt_1}{2\pi} e^{-i\omega_1t_1}\int e^{i\omega t}dt \sigma^{\mu\alpha}(t,t_1),
    \label{eq:sigma_wt}
\end{align}
where in the second line we convert the functional derivative with respect to the external field in the frequency domain to the time domain.

To evaluate $\sigma^{\mu\alpha}(t,t_1)$, we note that the velocity operator in the perturbed system also depends on the external field, cf. Eq. (\ref{eq:J}) and Eq. (\ref{eq:v_t}), so the functional derivatives can be taken on the observable $v^\mu(t)$ or on the exponent in $H_E(t)$. They are,
\begin{equation}
    \frac{\delta \hat{v}^{\mu}(t)}{\delta E^{\alpha}(t_{1})}-i\hat{v}^{\mu}(t)\frac{\delta}{\delta E^{\alpha}(t_{1})}\int dt'H_{E}(t').
    \label{eq:sigma_11+sigma_12}
\end{equation}
We observe that velocity operators, hence the electron-photon coupling vertex can come from either the current density operator or from $H_E$ in Eq. (\ref{eq:sigma_11+sigma_12}), which motivates the authors in Ref.~\cite{Parker2019} to define the outgoing vertex for the former and the incoming vertex for the latter, respectively.
After performing the functional derivative on Eq. (\ref{eq:v_t}) and Eq. (\ref{eq:H_E}), we have,
\begin{align*}
\sigma^{\mu\alpha}(t,t_{1}) & =i\text{e}^{2}\int\frac{d\omega_{1}}{2\pi}\,\frac{e^{-i\omega_{1}(t-t_{1})}}{\omega_{1}}\left\langle \hat{h}^{\mu\alpha}(t)\right\rangle \\
 & -i\text{e}^{2}\int dt'\int\frac{d\omega_{1}}{2\pi}\,\frac{e^{-i\omega_{1}(t'-t_{1})}}{\omega_{1}}\left\langle \hat{h}^{\mu}(t)\hat{h}^{\alpha}(t')\right\rangle .
\end{align*}

We proceed by writing the expectation values explicitly in terms of the matrix element $h^{\alpha}_{ab}$ and the two-particle correlation function. We focus on the second term, which dominates the finite frequency response for a semiconductor when including excitonic effects. 
We have within the IPA
\begin{align*}
&\expval{\hat{h}^{\mu}(t)\hat{h}^{\alpha}(t')}  
 =-h_{ab}^{\mu}h_{ba}^{\alpha}G_{b}(t,t')G_{a}(t',t)\nonumber\\
 & =-h_{ab}^{\mu}h_{ba}^{\alpha}\int\frac{d\omega''}{2\pi}e^{-i\omega''(t-t')}\int\frac{d\omega'}{2\pi}e^{-i\omega'(t'-t)} \nonumber\\
 &\times G_{b}(\omega'')G_{a}(\omega'),
\end{align*}
where the single particle Green's function is defined as $G_{a}(\vk; t-t^+)\equiv -\delta_{ab}\expval{c_{\vk a}(t)c^{\dagger}_{\vk b}(t^+)}$. It is important to note that the expectation value is taken at the unperturbed state. We adopt the convention that the repeated indices that do not show up on the left-hand side of the equation are summed.
Inserting the frequency integral of Green's function given in Appendix A, and using Eq. (\ref{eq:sigma_wt}), we obtain the first-order conductivity from the second term in Eq. (\ref{eq:sigma_11+sigma_12}), which describes the interband transition, 
\begin{equation}
    \sigma_{IP}^{\mu\alpha}(\omega;\omega_{1})=-\frac{iC_1}{\hbar\omega}\frac{f_{ab}h_{ba}^{\alpha}h_{ab}^{\mu}}{\hbar\omega_{1}-\varepsilon_{ba}+i\eta},
    \label{eq:sigmaIP}
\end{equation}
where $f_{ab}=f_{a}-f_{b}$ is defined as the difference of electron occupations between $a$ and $b$ bands, and $\varepsilon_{ba}=\varepsilon_{b}-\varepsilon_{a}$ is the difference of their energy. We define $C_1=g\text{e}^2\hbar/V_{tot}$, where the factor $g$ accounts for the spin degeneracy factor, and $V_{tot}=N_k V_u$ is the total volume with $N_k$ being the total number of $k$ points and $V_u$ being the unit cell volume and $\eta$ is a small positive number. This result agrees with the derivation from density matrix formulation~\cite{Sipe2000,Parker2019}.

Within the IPA, the first-order conductivity is given by the two diagrams,
\begin{multline}
    \sigma^{\mu\alpha}_{IP}(\omega;\omega_{1}) = \\
    \raisebox{-1cm}{\includegraphics[]{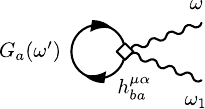}
}+\ \ 
    \raisebox{-1cm}{\includegraphics[]{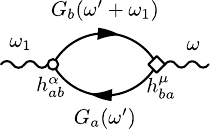}
}\\
\label{eq:1_IP_diagram}
\end{multline}
where $\omega_{1}$, $\omega$ indicates absorbed and emitted photons respectively. The first term corresponds to the Drude weight~\cite{Parker2019} and the second term describes the interband optical transitions. The bubble diagram in the second term represents the free electron-hole correlation function. Physically, the second diagram can be read as follows. An incoming photon generates a free electron-hole pair which later recombines and emits a photon. Expressions of components in the diagrams are given in Table I. We note that different symbols are used to distinguish incoming and outgoing photon vertices. In particular, the outgoing vertex associated with the current observable is represented by an empty diamond while the incoming vertex, which is associated with the perturbation expansion of $H_E(t)$, is represented by an empty circle. For the second order response, the conductivity tensor $\sigma^{\mu\alpha\beta}(\omega;\omega_{1},\omega_{2})$ can be computed from the second derivative of current expectation value, $\frac{\delta J(t)}{\delta E(t_{1})\delta E(t_{2})}$. A detailed derivation is given in Appendix~\ref{sec:appendixB1}.

\begin{table}[t]
\centering
\caption{Diagram components for the conductivity within IPA.}
\begin{tabular}[c]{ccc}
\hline\hline
 \multicolumn{1}{c}{\makecell[c]{Physical Description} } & Diagram   & \multicolumn{1}{c}{\makecell[c]{Mathematical\\ Expression}}  \\
\hline
 \multicolumn{1}{c}{\makecell[c]{Incoming vertex} } & 
 \begin{minipage}[b]{0.025\columnwidth}
    \centering
	\raisebox{-.5\height}{\includegraphics[width=\linewidth]{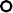}}
 \end{minipage} & $h$ \\
 \multicolumn{1}{c}{\makecell[c]{Outgoing vertex} } & 
 \begin{minipage}[b]{0.03\columnwidth}
    \centering
	\raisebox{-.5\height}{\includegraphics[width=\linewidth]{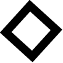}}
 \end{minipage} & $h$ \\
 \multicolumn{1}{c}{\makecell[c]{Incoming photon\\ with polarization $\alpha_{n}$} } & 
 \begin{minipage}[b]{0.15\columnwidth}
    \centering
	\raisebox{-.5\height}{\includegraphics[width=\linewidth]{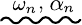}}
 \end{minipage} & $\omega_{n},\ \alpha_{n}$ \\
 \multicolumn{1}{c}{\makecell[c]{Outgoing photon \\ with polarization $\mu$} } & 
 \begin{minipage}[b]{0.15\columnwidth}
    \centering
	\raisebox{-.5\height}{\includegraphics[width=\linewidth]{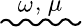}}
 \end{minipage} & $\omega,\ \mu$ \\
 \multicolumn{1}{c}{\makecell[c]{Multi-photon absorption} }& 
 \begin{minipage}[b]{0.3\columnwidth}
    \centering
	\raisebox{-.5\height}{\includegraphics[width=\linewidth]{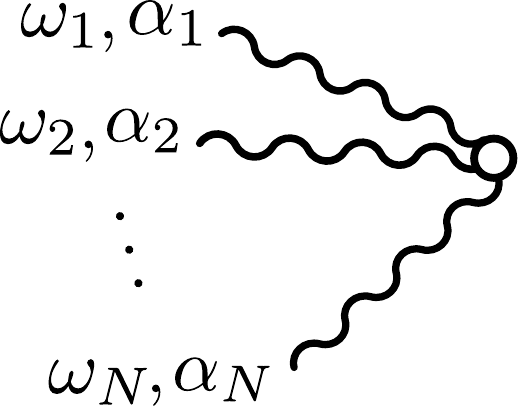}}
 \end{minipage}
 & $h^{\alpha_{1} \alpha_{2} \cdots \alpha_{N}}$ \\
 \multicolumn{1}{c}{\makecell[c]{Multi-photon absorption \\ and single photon emission} }  & 
 \begin{minipage}[b]{0.3\columnwidth}
    \centering
	\raisebox{-.5\height}{\includegraphics[width=\linewidth]{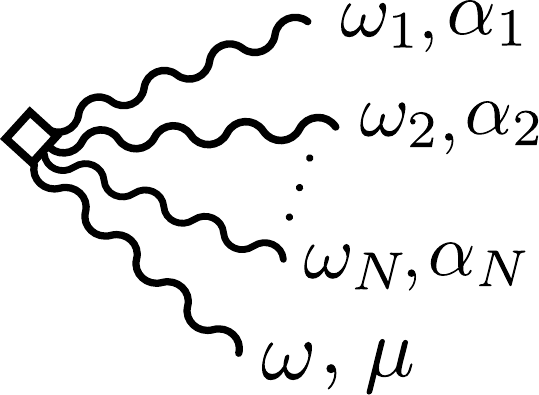}}
 \end{minipage}
 & $h^{\mu \alpha_{1} \alpha_{2} \cdots \alpha_{N}}$ \\
 \multicolumn{1}{c}{\makecell[c]{\\ Electron propagator \\} } & 
 \begin{minipage}[b]{0.25\columnwidth}
    \centering
	\raisebox{-.5\height}{\includegraphics[width=\linewidth]{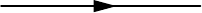}}
 \end{minipage}
 & $G$ \\
\hline
\end{tabular}
\end{table}

\section{Diagrammatic approach to excitonic effects}
\label{sec:eh_diagrams}
To incorporate excitonic effects into our derivation, we start with the interacting electron-hole correlation function, the equation of motion of which is described by the BSE~\cite{Rohlfing2000, Onida2002,stefanucci2013}. By taking the static approximation on the screened Coulomb interactions one can construct the correlation function from the eigensolutions of the BSE. To include excitonic effects into the derivation one can add all possible ladder diagrams, which describe repeated electron-hole interactions, into the IP conductivity diagrams. This can be done systematically by dressing the electron-photon vertex~\cite{Sven2020} as we show in the following. 

\subsection{Electron-hole interaction}
\label{sec:e-h interaction}
\begin{figure}[t]
\centering
    \begin{align*}
    \raisebox{-0.8cm}{\includegraphics[height=1.6cm]{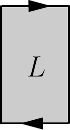}} &=
    \raisebox{-0.8cm}{\includegraphics[height=1.6cm]{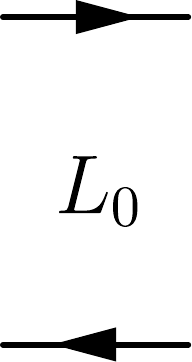}}+
    \raisebox{-0.8cm}{\includegraphics[height=1.6cm]{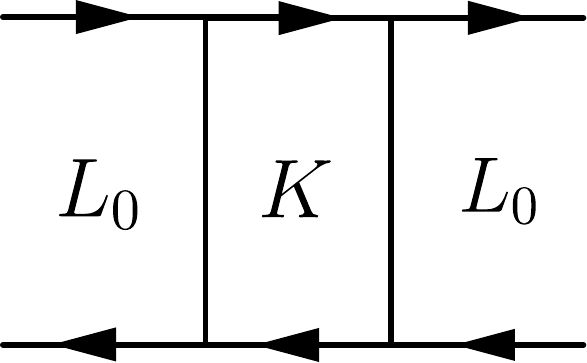}}  \\& \ \ \ \ \ +
    \raisebox{-0.8cm}{\includegraphics[height=1.6cm]{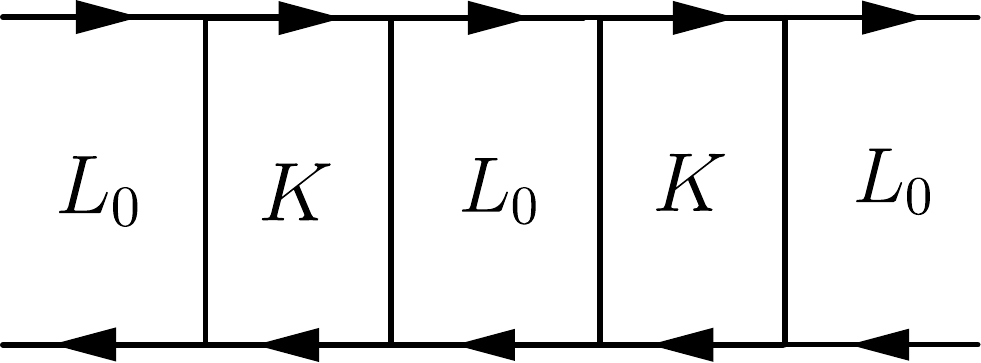}}+ \ldots
    \\& = 
    \raisebox{-0.9cm}{\includegraphics[height=1.6cm]{ladder2-2.pdf}}+
    \raisebox{-0.9cm}{\includegraphics[height=1.6cm]{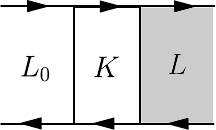}}\ \ 
    \end{align*}
    \begin{align*}
    \raisebox{-0.7cm}{\includegraphics[height=1.6cm]{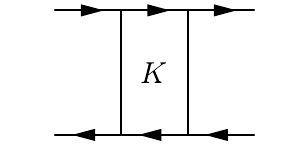}} =
    \raisebox{-0.7cm}{\includegraphics[height=1.6cm]{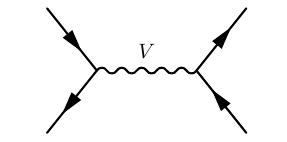}}-
    \raisebox{-0.9cm}{\includegraphics[height=2cm]{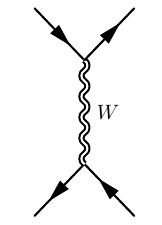}}
    \end{align*}
    \caption{Top panel: Diagrammatic representation of the Bethe-Salpeter equation for the electron-hole correlation function. The symbols $L$ and $L_{0}$ represent the electron-hole correlation function for interacting and independent particles, respectively. Bottom panel: $K$ denotes the interaction kernel which includes an attractive screened electron-hole interaction, $W$ (double-wiggly line), and a repulsive Coulomb exchange term, $V$ (single-wiggly line).}
    \label{fig:BSE_Kernel}
\end{figure}

Excitons are bound electron-hole pairs, where electrons and holes interact with each other via Coulomb interactions. 
For an interacting electron-hole correlation function $L$, its equation of motion is the BSE, 
\begin{multline}
    L_{\mathbf{k}ab,\mathbf{k}'cd}(\omega)=L_{0;\mathbf{k},ab}(\omega)\delta_{ac}\delta_{bd} \\ +L_{0;\mathbf{k},ab}(\omega)\sum_{\mathbf{k}'',l,l'}K_{\mathbf{k}ab,\mathbf{k}''ll'}L_{\mathbf{k}''ll',\mathbf{k}'cd}(\omega),
\label{eq:BSE_L}
\end{multline}
where we denote the non-interacting electron-hole correlation function as $L_{0;\mathbf{k},ab}$ with band indices in alphabet letters and the crystal momentum $\mathbf{k}$. The BSE kernel $K$ describes the interaction between electron-hole pairs. In the approximation consistent with the GW quasi-particle self-energy, the kernel $K$ includes an attractive screened Coulomb potential $W(\omega)$ and a repulsive bare exchange term $V$.
Eq. (\ref{eq:BSE_L}) is shown in terms of diagrams in Fig.~\ref{fig:BSE_Kernel}.

For a general BSE, the screened Coulomb interaction in the kernel $K$ is frequency dependent and the band indices run over all bands. In practice, a static approximation to the screened Coulomb interaction $W(\omega)=W(0)$ is often assumed so that the BSE can be transformed into an eigenvalue problem~\cite{Rohlfing2000,Onida2002}. Moreover, we focus on semiconductors and use the Tamm-Dancoff approximation. With these assumptions, we can write the eigenvalue equation for excitons with zero center-of-mass momentum (See Ref.~\cite{Strinati1988,Rohlfing2000,Onida2002,stefanucci2013} and Appendix~\ref{sec:appendixA}),
\begin{equation}
H^{BSE}_{cv\vk,c'v'\vk'}Y^{s}_{c'v'\vk'}=E_{s}Y^{s}_{cv\vk}, \nonumber
\label{eq:BSE}
\end{equation}
where the index $s$ labels exciton states, $E_s$ and $Y^{s}_{cv\vk}$ are the exciton energy and the exciton envelope function, respectively, and the effective Hamiltonian $H^{BSE}_{cv\vk,c'v'\vk'}$ is
\begin{equation}
    H^{BSE}_{cv\vk,c'v'\vk'} = (\varepsilon_{c\vk}-\varepsilon_{v\vk})\delta_{cc'}\delta_{vv'}\delta_{\vk\vk'}+K_{cv\vk,c'v'\vk'}, \nonumber
\label{eq:BSE_2}
\end{equation}
where $\varepsilon_{v\vk}$ and $\varepsilon_{c\vk}$ are the energy of the conduction and the valence band electrons, respectively.
The interacting electron-hole correlation function $L_{ij\mathbf{k},nm\mathbf{k}'}$ can be written in terms of exciton solutions as 
\begin{dmath}
    L_{ij\mathbf{k}nm\mathbf{k}'}(\omega)=i\sum_{\lambda}\left[\bar{f}_{i}f_{j}\bar{f}_{n}f_{m}\frac{Y_{ij\mathbf{k}}^{\lambda}Y_{nm\mathbf{k}'}^{\lambda*}}{\omega-E_{\lambda}+i\eta} \\
     -f_{i}\bar{f}_{j}\bar{f}_{m}f_{n}\frac{Y_{ji\mathbf{k}}^{\lambda*}Y_{mn\mathbf{k}'}^{\lambda}}{\omega+E_{\lambda}+i\eta} \right],
\label{eq:L}
\end{dmath}
where $\bar{f}_n=1-f_n$.
The first term in the parenthesis is the resonant part while the second term is the anti-resonant contribution \cite{Onida2002}.

\subsection{Vertex correction}
\begin{figure}[t]
\centering
    \begin{align*}
    \raisebox{-0.7cm}{\includegraphics[height=1.4cm]{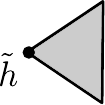}} &\ \ \ \ \ =\ \ \ \ \ 
    \raisebox{-0.5cm}{\includegraphics[height=0.6cm]{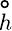}}\ \ \ \ \ +\ \ \ \ \ 
    \raisebox{-1.05cm}{\includegraphics[height=2.1cm]{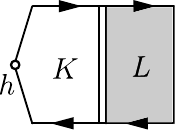}}
    \end{align*}
    \caption{The BSE for the dressed electron-photon vertex.}
    \label{fig:BSE_vertex}
\end{figure}
In Eq. (\ref{eq:sigmaIP}), we derive the interband IP conductivity by connecting the edges of electron and hole propagators of a free electron-hole correlation function with the velocity operators as shown in the second diagram in Eq. (\ref{eq:1_IP_diagram}).
To derive the optical conductivity tensor with excitonic effects, we are motivated to replace free electron-hole propagators with their interacting counterparts. 
The derivation of the first order optical conductivity can be done straightforwardly as shown in the next section.
For higher order responses, diagrams of three- or multi-particle correlation functions appear. 
In particular, the three-particle triangle diagram which represents the three-particle correlation function appears in the second order optical response.

Since a direct calculation of the interacting triangle diagram is challenging, inspired by Ref.~\cite{Sven2020} we approximate it with a series of ladder diagrams.
The essential idea is to add all possible non-crossing interaction lines for each pair of electron and hole legs.
We illustrate the procedure for the electron-photon vertex in Fig.~\ref{fig:BSE_vertex}.
The shaded triangle in Fig.~\ref{fig:BSE_vertex} stands for a dressed vertex, which can be expanded with respect to the order of interactions. 
At the zeroth order, we have a bare non-interacting vertex $\hat{h}$. In the first order, a single interaction kernel together with a free electron-hole correlation is included in the diagram.
At higher order, we insert more interaction kernels and $L$. The infinite order sum leads to the right-hand side of Fig.~\ref{fig:BSE_vertex}, where the infinite sum of ladder diagrams is replaced with the interacting electron-hole correlation. 
The BSE for the dressed incoming electron-photon vertex $\Tilde{h}$ satisfies~\cite{Sven2020}
\begin{equation}
    \Tilde{h}(\omega)=h+KL(\omega)h, \nonumber
\label{eq:BSE_h}
\end{equation}
where $h^{\mu}$ is the bare vertex.
Formally, in the matrix notation, the dressed incoming vertex can be solved by the inverse of the free electron-hole correlation,
\begin{equation}
    \Tilde{h}_{ab\mathbf{k}}^{\alpha}(\omega)=L_{0,ab\mathbf{k}}^{-1}(\omega)L(\omega)_{ab\mathbf{k}cd\mathbf{k}'}h_{cd\mathbf{k}'}^{\alpha}. \nonumber
\label{eq:h_measured}
\end{equation}
The dressed outgoing vertex is solved by,
\begin{equation}
    \Tilde{h}_{ab\mathbf{k}}^{\mu *}(\omega)=h_{cd\mathbf{k}'}^{\mu*}L(\omega)_{cd\mathbf{k}' ab\mathbf{k}}L_{0,ab\mathbf{k}}^{-1}(\omega). \nonumber
\label{h_incoming}
\end{equation}
\begin{figure}[t]
\centering
\begin{multline*}
    \raisebox{-1.25cm}{\includegraphics[height=2.5cm]{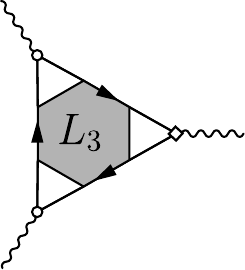}} \ \ \approx
    \raisebox{-1.25cm}{\includegraphics[height=2.5cm]{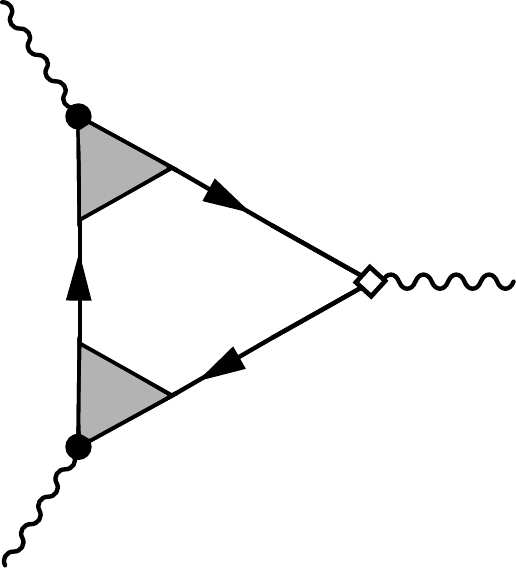}} \\ +
    \raisebox{-1.25cm}{\includegraphics[height=2.5cm]{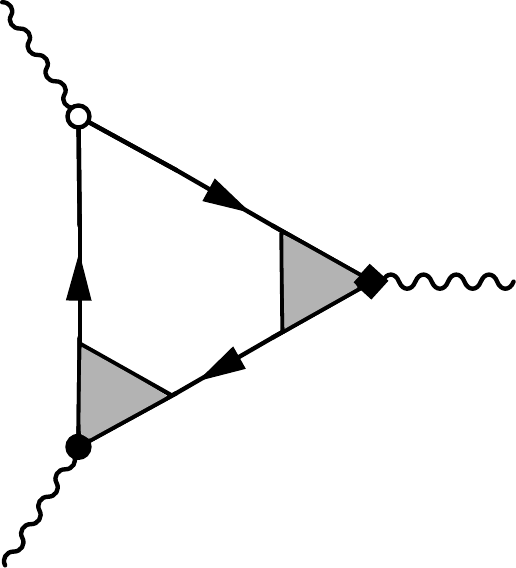}} +
    \raisebox{-1.25cm}{\includegraphics[height=2.5cm]{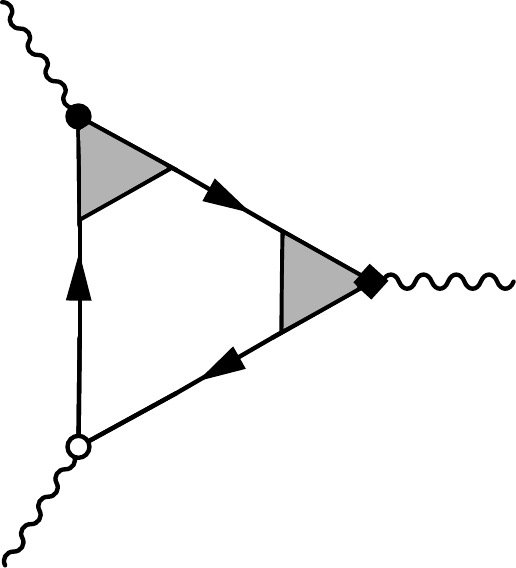}} ,
\end{multline*}
    \caption{The interacting three-particle correlation function diagram is approximated as a sum of three bare ones with two dressed vertices in each diagram.}
    \label{fig:3pt_diagram}
\end{figure}

The dressed vertex provides a systematical way to include excitonic effects in diagrams. For second order optical conductivity, within the ladder approximation, the interacting triangle diagram can be decomposed into a sum of three diagrams, each of which has two dressed vertices as shown in Fig.~\ref{fig:3pt_diagram}. 
We note that the diagram with all three vertices dressed simultaneously is not included since it would involve an electron-electron or hole-hole correlation function, which is beyond the Tamm-Dancoff approximation.

We conclude that excitonic effects can be included in the N-th order conductivity within the Tamm-Dancoff approximation by simply dressing all vertices in a way that,
\begin{itemize}
    \item Electron and hole propagators of any two e-h correlation functions associated with the dressed vertices can not directly connect to each other.
    \item By associating one of the two legs of an e-h correlation function with an electron and the other with a hole for every propagator in a diagram,  all legs can not have mixed characters of electron and hole.
\end{itemize}
The first rule is to avoid the double-counting while the second follows from the Tamm-Dancoff approximation.
New components with dressed vertices up to the second order response are listed in Table~\ref{table:eh}.

\begin{table}[t]
\centering
\caption{Diagram components for the conductivity tensor including excitonic effects within the ladder approximation.}
\begin{tabular}[c]{ccc}
\hline\hline
 \multicolumn{1}{c}{\makecell[c]{Physical Description } } & Diagram   & \multicolumn{1}{c}{\makecell[c]{Mathematical\\ Expression}}  \\
\hline
 \multicolumn{1}{c}{\makecell[c]{Dressed incoming\\ /outgoing vertex\\ \,}} & 
 \begin{minipage}[b]{0.08\columnwidth}
    \centering
	\raisebox{-.3\height}{\includegraphics[width=\linewidth]{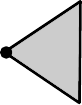}}
 \end{minipage}
 \begin{minipage}[b]{0.08\columnwidth}
    \centering
	\raisebox{-.3\height}{\includegraphics[width=\linewidth]{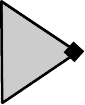}}
 \end{minipage}
 & $\Tilde{h}$\\
 \multicolumn{1}{c}{\makecell[c]{Interacting two-particle \\ correlation with \\ two bare vertices} }  & 
 \begin{minipage}[b]{0.25\columnwidth}
    \centering
	\raisebox{-.5\height}{\includegraphics[width=\linewidth]{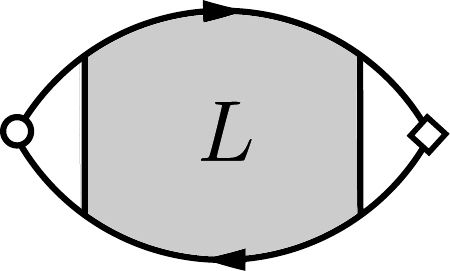}}
 \end{minipage}
 & $hLh$ \\
 \multicolumn{1}{c}{\makecell[c]{Bare two-particle \\ correlation with \\ the left vertex dressed} }  & 
 \begin{minipage}[b]{0.25\columnwidth}
    \centering
	\raisebox{-.5\height}{\includegraphics[width=\linewidth]{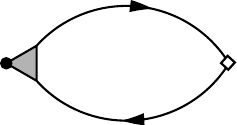}}
 \end{minipage}
 & $\Tilde{h}L_{0}h$ \\
 \multicolumn{1}{c}{\makecell[c]{Bare two-particle \\ correlation with \\ the right vertex dressed} }  & 
 \begin{minipage}[b]{0.25\columnwidth}
    \centering
	\raisebox{-.5\height}{\includegraphics[width=\linewidth]{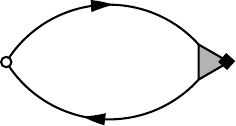}}
 \end{minipage}
 & $hL_{0}\Tilde{h}$ \\

 \multicolumn{1}{c}{\makecell[c]{One term of \\ the approximation to \\ the interacting \\ three-particle correlation} }  & \ \ \ 
 \begin{minipage}[b]{0.18\columnwidth}
    \centering
	\raisebox{-.5\height}{\includegraphics[width=\linewidth]{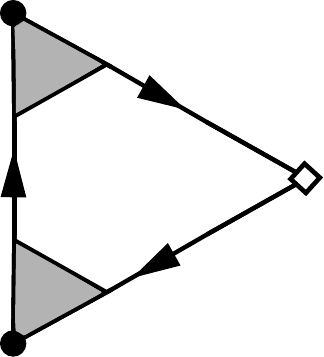}}
 \end{minipage}
 & $\Tilde{h}G\Tilde{h}GhG$ \\
\hline
\end{tabular}
\label{table:eh}
\end{table}

\section{Optical responses with excitonic effects}
\label{sec:optical_with_eh}
In this section, we derived expressions for optical conductivity up to the second order following the prescription introduced in Sec.~\ref{sec:e-h interaction}. 

\subsection{First order conductivity}
\label{sec:first_order}
\begin{figure}[tbh]
\centering
\begin{multline*}
\ \ \ \ \ \ \ \ 
    \raisebox{-0.7cm}{\includegraphics[height=1.5cm]{Feynman_diagrams/correlated_bubble.pdf}}\ \ \ =\ \ \ 
    \raisebox{-0.7cm}{\includegraphics[height=1.5cm]{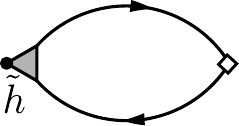}}
    \\ \ \ 
    \raisebox{-0.5cm}{\includegraphics[height=1.2cm]{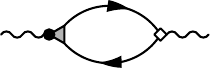}
    }\ = \ \ 
\raisebox{-0.5cm}{\includegraphics[height=1.2cm]{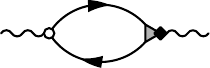}
}
\end{multline*}
    \caption{Top panel: the transformation of an interacting electron-hole bubble diagram into a bare bubble with a dressed vertex. Bottom panel: equivalence of dressing vertex on either side of the bubble.}
    \label{fig:shaded bubble}
\end{figure}

For semiconductors, we ignore the first diagram in Eq. (\ref{eq:1_IP_diagram}), which corresponds to the Drude weight.
The second term which describes the interband transition can be expressed in terms of bare vertices and the interacting two-particle correlation function.
Replacing the free e-h bubble with Eq. (\ref{eq:L}), the expression for the resonant part reads,
\begin{align}
\sigma_{eh}^{\mu\alpha}(\omega;\omega_{1}) & =\frac{iC_1}{\hbar\omega}\sum_{s}h_{ab}^{\mu,*}\frac{Y_{ab\mathbf{k}}^{s}Y_{ij\mathbf{k}'}^{s*}}{\hbar\omega_{1}-E_{s}+i\eta}h_{ij}^{\alpha}\nonumber \\
 & =\frac{iC_1}{\hbar\omega}\sum_{s}\frac{d_{s}^{\mu*}d_{s}^{\alpha}}{\hbar\omega_{1}-E_{s}+i\eta}. \nonumber
\end{align}
where we defined the excitonic velocity $d^{\alpha}_{s}$, 
\begin{equation}
    d_{s}^{\alpha}=Y_{ab\mathbf{k}}^{s*}h_{ab\mathbf{k}}^{\alpha},
\label{eq:d_s}
\end{equation}
where $\alpha$ is the polarization direction and $s$ is the exciton state index.
The real part of $\sigma^{\mu\alpha}(\omega;\omega)$ is related to the optical absorption spectrum, which is often represented by the imaginary part of the dielectric function $\epsilon_2(\omega)$.
We have $\epsilon_2(\omega)=\text{Re}[\sigma(\omega)]/(\epsilon_0\omega)$ so 
\begin{equation}
    \epsilon^{\mu\alpha}_2(\omega)=\frac{\pi\text{e}^2g}{\epsilon_0\omega^2 V_{tot}}\sum_s d_{s}^{\mu*}d_{s}^{\alpha} \delta(\hbar\omega-E_{s}),
\label{eq:absorption}
\end{equation}
where $\epsilon_0$ is the vacuum permittivity. Eq. (\ref{eq:absorption}) is the familiar expression for optical absorption spectrum in the literature~\cite{Deslippe2012}.

Equivalently, we can consider the excitonic effect by dressing one of the electron-photon vertex and evaluate the following diagram

\begin{multline}
\sigma^{\mu\alpha}_{eh}(\omega;\omega_{1}) = 
\raisebox{-1cm}{\includegraphics[]{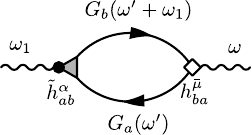}}, \nonumber
\end{multline}
where we replace the interacting bubble diagram with the bare one but introduce the dressed vertex $\Tilde{h}$. 
The mathematical expression for the above diagram is
\begin{multline}
    \sigma^{\mu\alpha}_{eh}(\omega;\omega_{1})=\frac{iC_1}{\hbar\omega}\Tilde{h}^{\alpha}_{ab}L_{0,ab\mathbf{k}}(\omega_1)h_{ab}^{\mu*}\\  
  =\frac{iC_1}{\hbar\omega}L_{0,ab\mathbf{k}}^{-1}(\omega_{1})L_{ab\mathbf{k},ij\mathbf{k}'}(\omega_{1})h_{ij}^{\alpha}h_{ab}^{\mu*}L_{0,ab\mathbf{k}}(\omega_{1}), \nonumber
\end{multline}
where in the second line we insert the dressed vertex from Eq. (\ref{eq:L}).
The equivalence of these two derivations can be expressed diagrammatically as shown in the top panel in Fig.~\ref{fig:shaded bubble}, which demonstrates that we can ``push'' the shaded area into one of the vertex.
We can also show that it is flexible to dress either the incoming photon vertex or the outgoing one.

\subsection{Second-order conductivity}
The topologically inequivalent diagrams for the second order optical conductivity including excitonic effects are shown in Eq. (\ref{eq:2nd_eh_diagram}).
Compared with diagrams for IP conductivity, the diagram with three photon line on the outgoing vertex is ignored since there are no corresponding corrections within the Tamm-Dancoff approximation and we are dealing with cold semiconductors.
The first two terms are bubble diagrams with an either incoming or outgoing two-photon line. 
Excitonic effects are included by dressing one vertex, which is similar to the treatment for the linear order diagram and is equivalent to replacing the free electron-hole correlation function with the interacting one as shown in Sec.~\ref{sec:first_order}.
At each vertex, the energy conservation of the photon and the electron-hole pair can be checked from the frequency argument of the single-particle Green functions. The indices $a$ and $b$ run over pairs between valence and conduction bands.

The interacting triangle diagram acquires different corrections from excitonic effects. 
Following the discussion in Sec.~\ref{sec:e-h interaction}, it is approximated by the sum of three diagrams, each of which has two dressed vertices.
Again, energy conservation can be readily checked at each vertex.
We observe that the bare vertex in each diagram couples the electron or the hole states of the two correlated electron-hole pairs associated with the two dressed vertices. Moreover, the bare vertex only couples conduction or valence electrons in the same manifold.

\begin{widetext}
\begin{multline}
    \sigma^{\mu\alpha\beta}_{eh}(\omega;\omega_{1},\omega_{2}) =\ \ 
    \raisebox{-1cm}{\includegraphics[]{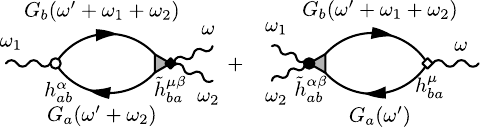}
    }\\+ \ \ \  
    \raisebox{-1.3cm}{\includegraphics[]{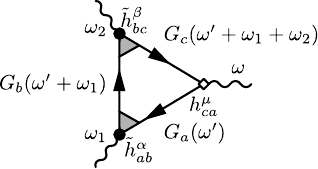}
    }+
  \raisebox{-1.4cm}{
    \includegraphics[]{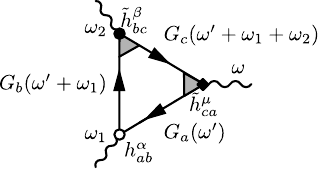}}\\+
\raisebox{-1.4cm}{
\includegraphics[]{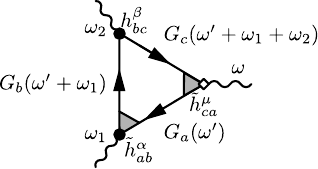}
} 
+ [(\alpha, \omega_{1})\leftrightarrow(\beta, \omega_{2})].
\label{eq:2nd_eh_diagram}
\end{multline}
\end{widetext}

After some tedious but straightforward algebra, we obtain an expression for the general second order optical response including excitonic effects. 
The details will be given in the Appendix~\ref{sec:appendixB1}. 
We write down the final expression, which roughly corresponds to the five diagrams above up to the symmetrization over photon frequency and polarizations,
\begin{widetext}
    \begin{equation}
    \begin{split}
  \sigma^{\mu\alpha\beta}_{eh}(\omega;\omega_{1},\omega_{2})\\
  = \ \ &\frac{-C_2}{\hbar^{2}\omega_{1}\omega_{2}}\sum_{\lambda}\left[\frac{d_{\lambda}^{\alpha}d_{\lambda}^{\mu\beta*}}{\hbar\omega_{1}-E_{\lambda}+i\eta}-\frac{d_{\lambda}^{\alpha*}d_{\lambda}^{\mu\beta}}{\hbar\omega_{1}+E_{\lambda}+i\eta}\right]+\frac{-C_2}{2\hbar^{2}\omega_{1}\omega_{2}}\sum_{\lambda}\left[\frac{d_{\lambda}^{\alpha\beta}d_{\lambda}^{\mu*}}{\hbar\omega-E_{\lambda}+i\eta}-\frac{d_{\lambda}^{\alpha\beta*}d_{\lambda}^{\mu}}{\hbar\omega+E_{\lambda}+i\eta}\right]\\
 & +\frac{C_2}{\hbar^{2}\omega_{1}\omega_{2}}\sum_{s\lambda}\left[\frac{d_{s}^{\alpha}\Pi_{s\lambda}^{\mu*}d_{\lambda}^{\beta*}}{(\hbar\omega_{2}+E_{\lambda}+i\eta)(\hbar\omega_{1}-E_{s}+i\eta)}-\frac{d_{\lambda}^{\mu*}\Pi_{\lambda s}^{\alpha}d_{s}^{\beta}}{(\hbar\omega-E_{\lambda}+i\eta)(\hbar\omega_{2}-E_{s}+i\eta)}\right]\\
 & -\frac{C_2}{\hbar^{2}\omega_{1}\omega_{2}}\sum_{s\lambda}\frac{d_{\lambda}^{\mu}\Pi_{s\lambda}^{\alpha}d_{s}^{\beta*}}{(\hbar\omega+E_{\lambda}+i\eta)(\hbar\omega_{2}+E_{s}+i\eta)} + [(\alpha, \omega_{1})\leftrightarrow(\beta, \omega_{2})],
\end{split}
\label{eq:2nd_eh}
\end{equation}
\end{widetext}
where we define the second order excitonic velocity matrix element, $d_\lambda^{\alpha \beta}=h_{ab}^{\alpha\beta}Y^{\lambda *}_{ab\mathbf{k}}$, the inter-exciton coupling matrix elements,
\begin{equation}
    \Pi_{\lambda s}^{\beta}=h_{cb}^{\beta}Y_{ca\mathbf{k}}^{\lambda*}Y_{ba\mathbf{k}}^{s}-h_{ba}^{\beta}Y_{ca\mathbf{k}}^{\lambda*}Y_{cb\mathbf{k}}^{s},
\label{eq:inter-exciton coupling}
\end{equation}
and $C_2=g\text{e}^3\hbar^2/V_{tot}$.
The first term in $\Pi_{\lambda s}$ describes the coupling between electrons in two excitons while the second term is for the coupling between holes. The inter-exciton coupling term originates from the bare vertex that appears in the triangle diagram.

Eq. (\ref{eq:2nd_eh}) is the main result of this work. The physical meaning of each term can be read out directly with the help of diagrams. Terms in the first parenthesis in Eq. (\ref{eq:2nd_eh}) correspond to the first diagram where an incoming photon created an electron-hole pair with energy $E_\lambda$, which later absorbed and at the same time emitted a photon through the second order velocity operator. The second term in the first parenthesis describes an anti-resonance process. The first term in the second parenthesis describes the excitation of an electron-hole pair by absorbing two photons, which emit a photon later. The 1/2 factor comes from the fact that this diagram originates from the second order expansion of the coupling Hamiltonian in the functional integral and an exchange of $\alpha\leftrightarrow\beta$ and $\omega_1\leftrightarrow\omega_2$ gives the same result.

The last three terms come from symmetrization and rearrangement of the three triangle diagrams.
We can interpret the third diagram in Eq. (\ref{eq:2nd_eh_diagram}) by starting from the bottom-left corner. An electron-hole pair was generated by absorbing a photon. The hole state was scattered to another hole state and emitted part of its energy at the bottom-right corner. The electron and the hole absorbed light but deexcited at the vertex on the top-left corner.
In the fourth diagram, an electron-hole pair is generated by absorbing a photon on the top-left corner. The hole is scattered by another photon in the bottom-left corner. The pair recombines and emits a photon with frequency $\omega$ at the right vertex.
In the fifth diagram, which has a correspondence to the symmetrized partner term in the last line in Eq. (\ref{eq:2nd_eh}), an anti-resonance process generates an electron-hole pair then the hole state is scattered at the bare vertex. Finally, the e-h pair recombines and emits a photon.

It is interesting to compare Eq. (\ref{eq:2nd_eh}) with other reported derivations. A comparison between Eq. (\ref{eq:2nd_eh}) and  Eq. (B1c) in Ref.~\cite{Taghizadeh2018} will be given in the next section. 
We will show that there is a one-to-one correspondence for each term between that expression and Eq. (\ref{eq:2nd_eh}). However, the two derivations differ in how the velocity operators are defined. 
In the next section, we will present the numerical results of the second harmonic generation (SHG) for a simple two-band model.

\section{Application on a monolayer hexagonal Boron-nitride}
\label{sec:model}
In this section, we apply our method to compute the linear absorption and the SHG spectrum for a model of monolayer hexagonal Boron nitride (h-BN). Since monolayer h-BN is a large band gap semiconductor it is known to have strong excitonic effects as we shall also demonstrate later.

\subsection{Two band tight-binding model}
We employ a two-band tight-binding model for monolayer h-BN~\cite{Sipe235307}. In the local basis of B and N atoms, the tight-binding Hamiltonian in the momentum space reads, 
\begin{equation*}
    H^{\text{hBN}}_{\mathbf{k}} = \begin{pmatrix}
        \Delta & t_{0}f_{\mathbf{k}} \\
        t_{0}f_{\mathbf{k}}^{*} & -\Delta
    \end{pmatrix},
\end{equation*}
where we define the structure factor \(f_{\mathbf{k}} = 1 + e^{-i\mathbf{k}\cdot\mathbf{a}_1} + e^{-i\mathbf{k}\cdot\mathbf{a}_2}\) with the primitive lattice vectors $\boldsymbol{a}_{1}=a_{0}\left(\frac{\sqrt{3}}{2}\hat{x}-\frac{1}{2}\hat{y}\right)$, $\boldsymbol{a}_{2}=a_{0}\left(\frac{\sqrt{3}}{2}\hat{x}+\frac{1}{2}\hat{y}\right)$, and  the lattice constant $a_{0}=2.46 \ \rm \AA$. The asymmetric on-site energies are denoted as $\Delta$ and $-\Delta$ for B and N atoms, respectively. We choose \(\Delta = 3.9\) eV and the nearest neighbor hopping strength \(t_0 = 2.7\) eV in our calculations below. 
The eigenenergies and eigenstates read,
\begin{equation}
    \begin{split}
        &\varepsilon_{c/v\mathbf{k}}=\pm \sqrt{\Delta^{2}+(t_{0}|f_{\mathbf{k}}|)^{2}} \\&
        \ket{c\vk}=\frac{1}{\sqrt{2}}
        \begin{pmatrix}
            \sqrt{1+\frac{\Delta}{\varepsilon_{c\mathbf{k}}}} \\
            \sqrt{1-\frac{\Delta}{\varepsilon_{c\mathbf{k}}}}\frac{f^{*}_{\mathbf{k}}}{\abs{f_{\mathbf{k}}}}
        \end{pmatrix} \\&
        \ket{v\vk}=\frac{1}{\sqrt{2}}
        \begin{pmatrix}
            -\sqrt{1-\frac{\Delta}{\varepsilon_{c\mathbf{k}}}}\frac{f_{\mathbf{k}}}{\abs{f_{\mathbf{k}}}} \\
            \sqrt{1+\frac{\Delta}{\varepsilon_{c\mathbf{k}}}}
        \end{pmatrix}.
    \end{split}
    \label{eq:TBsols}
\end{equation}
With our choices of parameters, the band gap is 2$\Delta$=7.8 eV.

Berry connection plays a crucial role in calculating optical matrix elements in both length gauge and velocity gauge. It is defined as \(\boldsymbol{\xi}_{nm\mathbf{k}} = iA^{-1}\int_A d\mathbf{r} u^*_{n\mathbf{k}}\boldsymbol{\nabla}_{\mathbf{k}}u_{m\mathbf{k}}\), where $A$ is the unit cell area and $u_{n\mathbf{k}}$ is the cell-periodic part of the Bloch state which is distinct from the eigenstates in Eq. (\ref{eq:TBsols}).
Following Ref.~\cite{Sipe235307}, we compute 
\begin{equation*}
\boldsymbol{\xi}_{nm\mathbf{k}}=i(\boldsymbol{\nabla}_{\mathbf{q}}U_{n\mathbf{k};m\mathbf{k}+\mathbf{q}})|_{\mathbf{q}=0},
\end{equation*}
where the overlap matrix element $U$ is defined as
\begin{equation*}            
U_{n\mathbf{k};m\mathbf{k}+\mathbf{q}}=\sum_{j=A,B}\braket{n\vk|j}\braket{j|m\vk+\mathbf{q}}e^{-i\mathbf{q}\cdot\boldsymbol{\tau_{j}}},
\end{equation*}
where $\braket{n\vk|j}$ is the wavefunction coefficients of the $j$-th sublattice orbital and the atom positions are given by $\boldsymbol{\tau}_{A}=\boldsymbol{0}$ and $\boldsymbol{\tau}_{B}=\frac{1}{3}(\boldsymbol{a}_{1}+\boldsymbol{a}_{2})$ for $A$ and $B$ sublattice, respectively.
The off-diagonal part of $\boldsymbol{\xi}_{nm\mathbf{k}}$ is the matrix elements of the position operator which describes interband optical transition amplitude. The Hermiticity of the overlap matrix can be seen from the definition.

We numerically solve BSE for excitons in this model. In our implementation, the repulsive exchange term $V$ is ignored and the attractive screened Coulomb interaction $W$ is modeled by a Yukawa form \cite{235432}
\begin{equation*}
    W_{vc,\mathbf{k}\mathbf{k}'}=-\frac{1}{8\pi^{2}\varepsilon\epsilon_{0}}
U_{v\mathbf{k}',v\mathbf{k}}U_{c\mathbf{k},c\mathbf{k}'}
\frac{e^{-l|{\mathbf{k}-\mathbf{k}'}|}}{|\mathbf{k}-\mathbf{k}'|},
\end{equation*}
where $U_{n\mathbf{k},m\mathbf{k}'}$ is the overlap matrix elements defined above, we set the effective thickness $l$ to be 1 $\rm \AA$ and the dielectric constant $\varepsilon=1.5$. The unit cell volume $V=Al$ is used in the calculation. Excitonic velocity matrix elements, $d^{\mu}_{s}$ from Eq.~(\ref{eq:d_s}), and the inter-exciton coupling matrix element $\Pi^{\alpha}_{\lambda s}$ defined in Eq.~(\ref{eq:inter-exciton coupling}) are then computed from exciton envelope functions for optical responses. The contraction over band indices is trivial for a two-band model.

\subsection{Optical response from the h-BN model}
\begin{figure}[t]
\centering
\includegraphics[width=8cm]{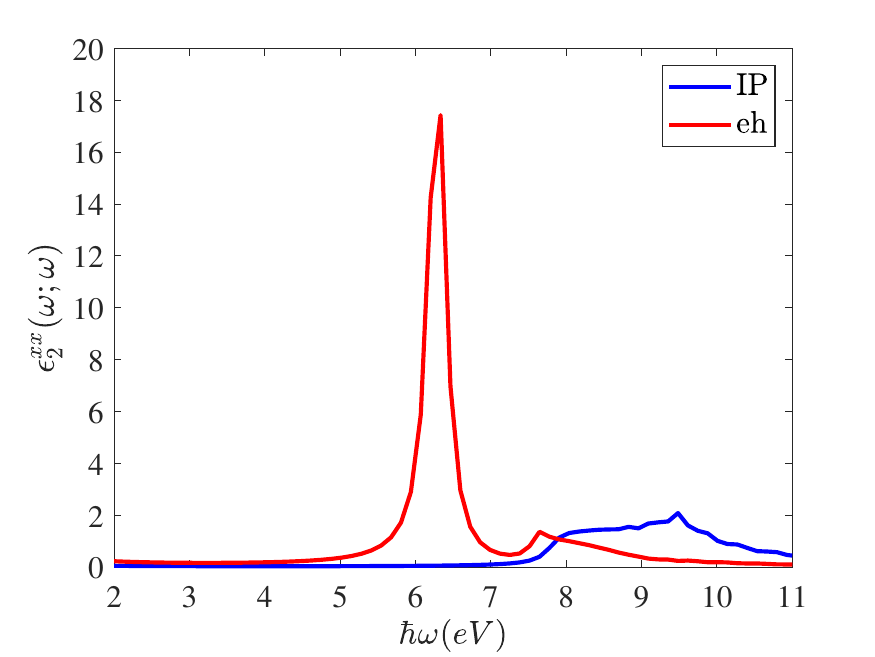}
\caption{The $xx$ tensor component of the imaginary part of the dielectric function computed without (blue solid line) and with (red solid line) excitonic effects.} 
\label{fig:1st}
\end{figure}

\begin{figure}[t]
\centering
\includegraphics[width=\linewidth]{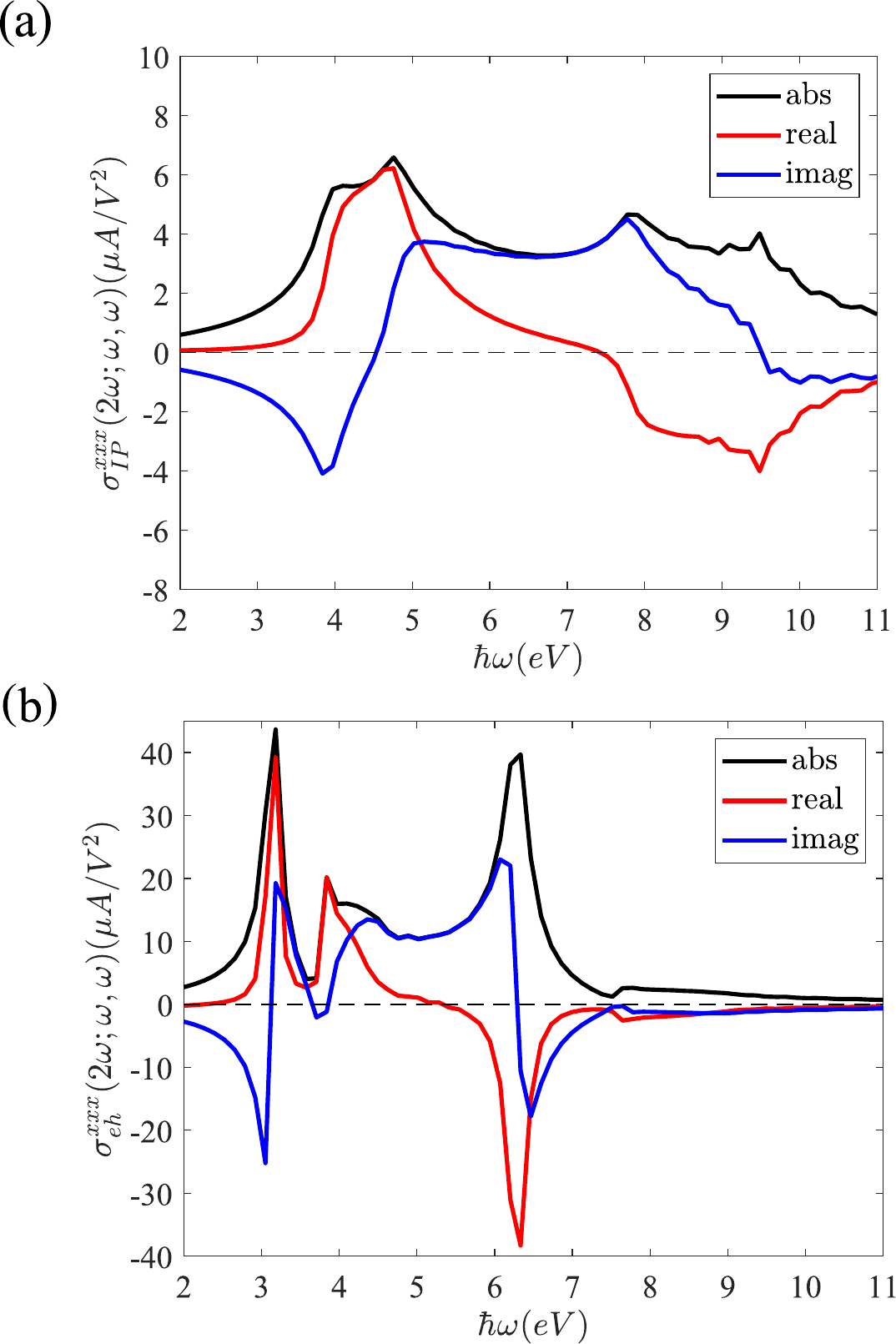}
 \caption{SHG conductivity spectrum, $\sigma_{eh}^{xxx}(2\omega;\omega, \omega)$, (a) computed with IPA and (b) computed from Eq.~\ref{eq:2nd_eh} including excitonic effects. Black, red, and blue solid lines are for the absolute value, the real part, and the imaginary part of the conductivity tensor, respectively.} 
\label{fig:SHG}
\end{figure}

We show the numerical results of absorption and SHG tensors for the two-band Hamiltonian in Fig.~\ref{fig:1st} and Fig.~\ref{fig:SHG}, respectively. Excitonic effects on these optical responses are demonstrated by comparing the IP results with those including e-h interactions.
In the IP case, we use the expression given in Ref.~\cite{235446}, which is derived in the velocity gauge within the density matrix formalism.

For the SHG conductivity tensor, we evaluate the expressions by setting $\omega_{1}=\omega_{2}=\omega$.
Lattice symmetry constrains that the $xxx$ tensor component is the only independent tensor component for second order conductivity tensors. Other components are either equal or differ by a minus sign.
Hence, we only show the $xxx$ component below. In our numerical calculations, a uniform $48 \times 48$ $\mathbf{k}$-grid and a broadening of 0.01 Ry is used. With the chosen parameters, we obtain a binding energy of 1.4 eV for the lowest exciton state.
The imaginary part of the dielectric function, which represents the absorption spectrum is shown in Fig.~\ref{fig:1st}. 
For IP results, We see a step function like a spectrum close to the band edge and a peak at 9.5 eV, which reflects the joint density of state of the monolayer h-BN in the same energy range.
Excitonic effects qualitatively change the spectrum. We observe that the spectrum manifests two peaks compared to the IP case, which can be attributed to A and B exciton excitations at 6.4 and 7.8 eV, respectively~\cite{Huang2023}. This result suggests substantial enhancement of the response due to excitonic effects. Our results generally agree with the reported results from model calculations~\cite{235432}.
We also confirm numerically that the results of Eq.~\ref{eq:2nd_eh} reduced to the IP results if we set the kernel to zero when solving the BSE Hamiltonian.

In Fig.~\ref{fig:SHG}, we show the $xxx$ tensor component of SHG conductivity, which are comparable to the results in Ref.~\cite{235432}. For the IP response shown in Fig.~\ref{fig:SHG} (a), higher responses are observed between 4-5 eV and 8-10 eV. The real part of the SHG responses between 8-10 eV resembles those in the absorption spectrum, which are due to the single photon resonance structure in the denominator of Eq.~\ref{eq:2nd_eh} while those between 4-5 eV can be understood as the corresponding two-photon resonance contribution.

Comparing Fig.~\ref{fig:SHG} (a) with (b), we can see that excitonic effects greatly enhance the SHG intensity and shift the spectrum to the low energy side. Moreover, two pronounced peaks appear; one is at 6.4 eV, which is identical to that in the absorption spectrum, and the second peak is located at 3.2 eV, which is half the energy of the first peak. The presence of these peaks is the consequence of either a one-photon or a two-photon resonance to the A exciton from the pole structure shown in Eq.~\ref{eq:2nd_eh}.
Secondary peaks due to B excitons can also be identified at 7.8 eV and 3.9 eV.
A detailed analysis reveals that the excitonic enhancement is a result from strong excitonic dipole matrix elements and large inter-exciton couplings between A and B excitons~\cite{Ruan2023}.

\subsection{Comparison with the density matrix formulation}
We compare our expression Eq.~\ref{eq:2nd_eh} with those derived from the density matrix formulation.
In Ref.~\cite{Taghizadeh2018}, expressions of second order conductivity tensors including excitonic effects in different gauges are derived from the density matrix formulation. 
In particular, Eq. (B1c) in Ref.~\cite{Taghizadeh2018}, which is also derived in the velocity gauge reads,
\begin{dmath}
    \sigma_{eh}^{\mu \alpha \beta}[\boldsymbol{\pi}_{\lambda s}](\omega;\omega_{1},\omega_{2})=
    \frac{C_2}{i\hbar^2\omega_{1}\omega_{2}}\sum_{\lambda} \left[
    \frac{A^{\mu \beta}
    _{\lambda}\pi^{\alpha *}_{\lambda}}{\hbar\omega_{1}-E_{\lambda}}+\frac{A^{\mu \beta *}
    _{\lambda}\pi^{\alpha}_{\lambda}}{\hbar\omega_{1}+E_{\lambda}}\right] \\
    -\frac{C_2}{2i\hbar^2\omega_{1}\omega_{2}}\sum_{\lambda} \left[
    \frac{\pi^{\mu}_{\lambda}A
    _{\lambda}^{\alpha \beta *}}{\hbar\omega-E_{\lambda}}+\frac{\pi_{\lambda}^{\mu *}A^{\alpha \beta}
    _{\lambda}}{\hbar\omega+E_{\lambda}} \right] \\
    +\frac{C_2}{\hbar^2\omega_{1}\omega_{2}} 
        \sum_{\lambda s} \left[
    \frac{\pi^{\mu}_{\lambda}\pi^{\alpha}
    _{\lambda s}\pi^{\beta *}_{s}}{(\hbar\omega-E_{\lambda})(\hbar\omega_{2}-E_{s})} \\+ 
    \frac{\pi^{\mu *}_{\lambda}\pi^{\alpha*}
    _{\lambda s}\pi^{\beta}_{s}}{(\hbar\omega+E_{\lambda})(\hbar\omega_{2}+E_{s})}-
    \frac{\pi^{\beta}_{\lambda}\pi^{\mu}
    _{\lambda s}\pi_{s}^{\alpha*}}{(\hbar\omega_{2}+E_{\lambda})(\hbar\omega_{1}-E_{s})} \right],
    \label{eq:VG1}
\end{dmath}
where $C_2$ is defined earlier and $\boldsymbol{\pi}_{\lambda}$, $\boldsymbol{\pi}_{\lambda s}$, and $\boldsymbol{A}_s$ are the excitonic velocity matrix elements, the inter-exciton velocity matrix elements, and the second order velocity matrix elements, respectively. They are defined as follows. 
We first define the excitonic dipole matrix element, $\boldsymbol{\xi}_\lambda=Y^{\lambda}_{cv\mathbf{k}}\boldsymbol{\xi}_{vc\mathbf{k}}$.
The definition of the matrix element of the excitonic velocity operator follows from the commutator between the BSE Hamiltonian and the excitonic dipole operator, $\boldsymbol{\pi}_{\lambda}=-iE_{\lambda}\boldsymbol{\xi}_{\lambda}$. 
Similarly, the inter-exciton velocity operator $\boldsymbol{\pi}_{\lambda s}$ is defined as, 
\begin{equation}
    \boldsymbol{\pi}_{\lambda s} = i(E_{\lambda}-E_{s})\boldsymbol{\xi}_{\lambda s}.
\label{eq:pi_s}
\end{equation}
The inter-exciton dipole matrix element $\boldsymbol{\xi}_{\lambda s}$ consists of inter- and intra-band part,
\begin{equation*}
        \boldsymbol{\xi}_{\lambda s}=\boldsymbol{Q}_{\lambda s} + \boldsymbol{R}_{\lambda s},    
\end{equation*}
where $\boldsymbol{R}_{\lambda s}$ represents the inter-band contribution within electrons or holes states in excitons, 
\begin{equation}
    \boldsymbol{R}_{\lambda s} = \sum_{cv\mathbf{k}}Y^{\lambda,*}_{cv\mathbf{k}}(\sum_{c_{1}\neq c}Y^{s}_{c_{1}v\mathbf{k}}\boldsymbol{\xi}_{cc_{1}\mathbf{k}}-
        \sum_{v_{1}\neq v}Y^{s}_{cv_{1}\mathbf{k}}\boldsymbol{\xi}_{v_{1}v\mathbf{k}})
\label{eq:R}
\end{equation}
while $\boldsymbol{Q}_{\lambda s}$ corresponds to the intra-band contribution described by the covariant derivative defined earlier,
\begin{equation*}
    \boldsymbol{Q}_{\lambda s}=i\sum_{\mathbf{k}}Y^{\lambda,*}_{cv\mathbf{k}}\boldsymbol{D}_g(Y^{s}_{cv\mathbf{k}}),
\end{equation*}
where $D^\alpha_g$ is the generalized derivative, defined as $D_g^{\alpha}O_{ab} = \frac{\partial O_{ab}}{\partial k^{\alpha}}-i(\xi_{aa}^{\alpha}-\xi^{\alpha}_{bb})O_{ab}$.
In a two-band model, $\boldsymbol{R}_{\lambda s}$ vanishes so we have $\boldsymbol{\xi}_{\lambda s}=\boldsymbol{Q}_{\lambda s}$.
The matrix element of the second-order velocity operator $A^{\alpha\beta}_{s}$ is defined by taking the commutator of the excitonic dipole operator and $\boldsymbol{\pi}_s$, which is expressed as $\boldsymbol{A}_{s} = \sum_{\lambda}(\boldsymbol{\xi}_{\lambda}\boldsymbol{\pi}_{\lambda s}-\boldsymbol{\pi}_{\lambda}\boldsymbol{\xi}_{\lambda s})$.

\begin{figure}[t]
    \centering
    \includegraphics[width=\linewidth]{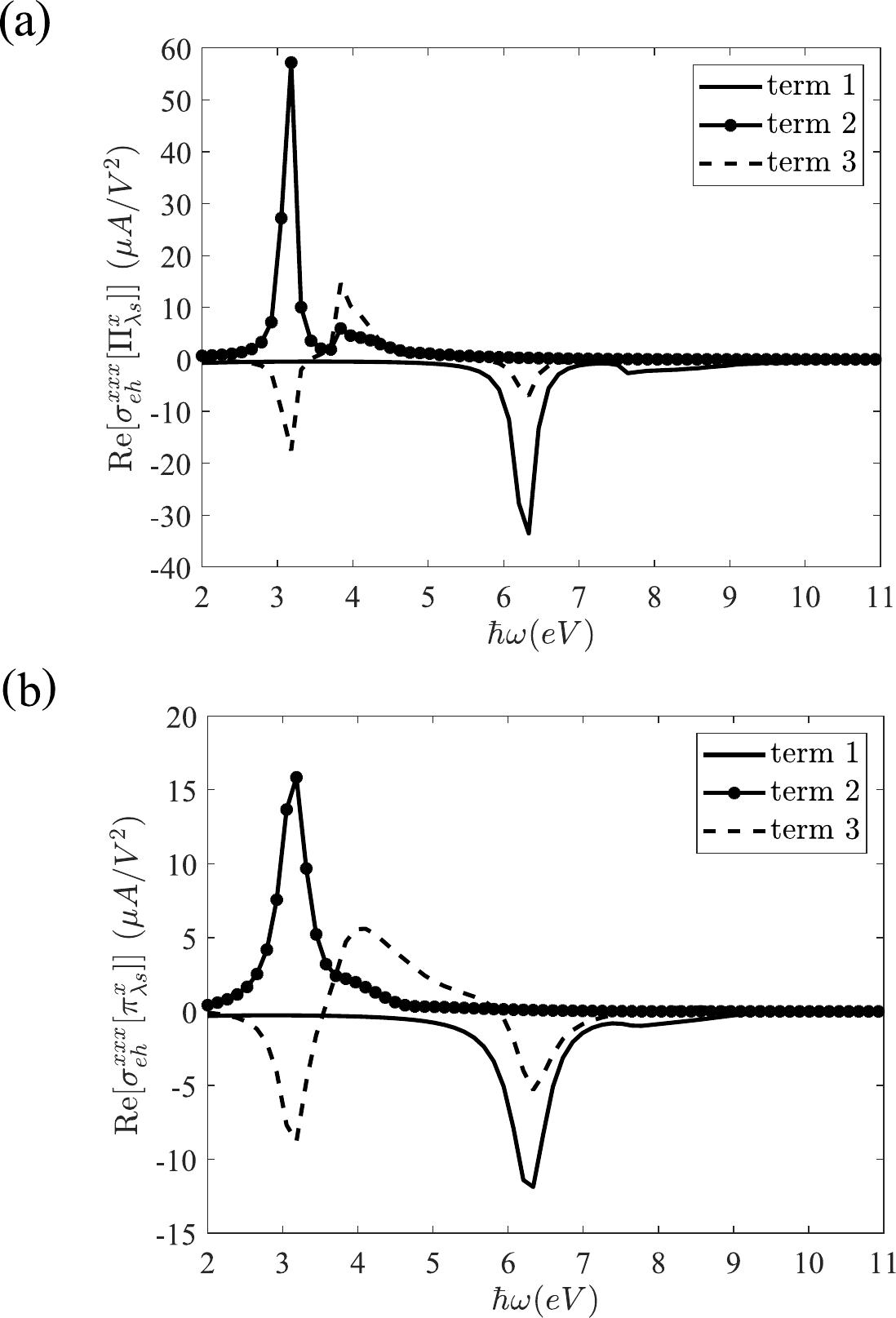}
    \caption{The real part of the three dominant terms to SHG. They correspond to the first term through the third term in (a) Eq. (\ref{eq:2nd_eh}) and (b) Eq. (\ref{eq:VG1}).}
    \label{fig:terms-SHG}
\end{figure}

In contrast to our excitonic operators defined earlier, the excitonic velocity operators here are defined through the commutator between the excitonic dipole operators and the exciton Hamiltonian.
By comparing Eq.~\ref{eq:2nd_eh} and Eq.~\ref{eq:VG1}, we can see that the two expressions can be related to each other by the following substitutions, $d^{\alpha}_{\lambda} \rightarrow \pi^{\alpha *}_{\lambda}$, $d^{\alpha \beta}_{\lambda} \rightarrow iA^{\alpha \beta *}_{\lambda}$, and $\Pi^{\alpha}_{\lambda s} \rightarrow \pi^{\alpha}_{\lambda s}$. For the correspondence between $\Pi_{\lambda s}$ and $\pi_{\lambda s}$, we note the similarity between Eq.~\ref{eq:inter-exciton coupling} and Eq.~\ref{eq:R}. The IP velocity matrix elements is related to the dipole matrix elements through Eq.~(\ref{eq:h_ab}).
In our definition, the excitonic velocity matrix elements are defined by contracting the IP velocity matrix elements with exciton envelope functions while it is defined as the two-particle generalization of the velocity operator in Ref.~\cite{Taghizadeh2018}.
A similar comparison also applies to the inter-exciton velocity and the second order excitonic velocity operators.
Such difference originates from the fact that we start with a single particle formulation and treat electron-hole interactions as a perturbation, while in Ref.~\cite{Taghizadeh2018} the derivation starts with the single-particle density matrix formulation but is generalized to excitonic operators defined from the two-particle Hamiltonian in the end.
\begin{figure}[t]
    \centering
    \includegraphics[width=\linewidth]{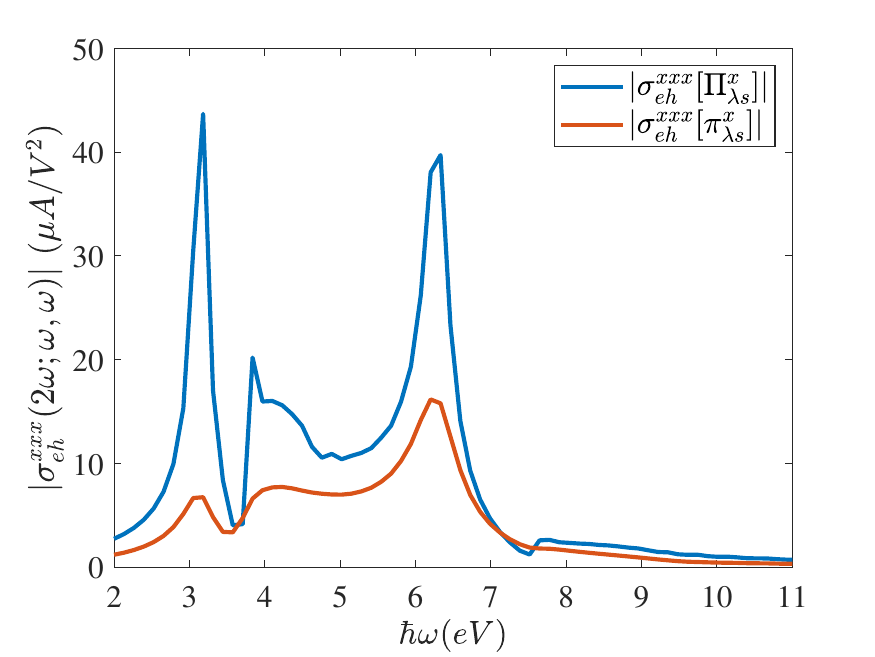}
    \caption{The absolute value of SHG conductivity tensors computed with Eq.~\ref{eq:2nd_eh} (blue solid line), $\sigma^{xxx}_{eh}[\Pi^x_{\lambda s}]$ and Eq.~\ref{eq:VG1} (orange solid line), $\sigma^{xxx}_{eh}[\pi^x_{\lambda s}]$.}
    \label{fig:gauges}
\end{figure}
In Fig.~\ref{fig:gauges}, we compare the numerical results from the two expressions. 
We observe that although the two results are qualitatively similar to each other, their absolute intensity differs. Our expression tends to give a higher conductivity than the result from Eq.~\ref{eq:VG1} and the first peak around 3 eV is more pronounced.
To further understand this difference, we analyze the contribution of each term separated by parenthesis in Eq.~\ref{eq:2nd_eh}.
We find that the first three terms have the dominant contributions to the SHG conductivity tensor. Their contributions are shown in Fig.~\ref{fig:terms-SHG} (a) and (b) for Eq.~\ref{eq:2nd_eh} and Eq.~\ref{eq:VG1}, respectivley.
From the peak position in each panel, we can identify their single or two-photon resonance origins.
A term-by-term comparison shows that the frequency dependence and the sign of corresponding terms from the two derivations qualitatively agree. 
However, the difference in their relative magnitude leads to the difference in total responses.
Specifically, the large deviation at around 3.0 eV is due to the cancellation between the contribution of the second and the third term as shown in Fig.~\ref{fig:terms-SHG} (b).
In contrast, the cancellation is less effective from our expression as shown in Fig.~\ref{fig:terms-SHG} (a).
We further compare the different matrix elements defined above.
For the inter-exciton coupling matrices, by definition the diagonal elements of $\pi^{\alpha}_{\lambda s}$ vanish while $\Pi^\alpha_{\lambda s}$ can have finite diagonal elements.
We find that the off-diagonal elements of $\Pi^{\alpha}_{\lambda s}$ is larger than $\pi^{\alpha}_{\lambda s}$ in magnitude in our calculation.
The comparison of the second order velocity matrix elements also shows a similar trend.

We argue that a possible reason for the discrepancy of the two derivations is the lack of self-consistency in our work. Since we use the bare Green function in all calculations, accordingly the velocity operator is defined from the corresponding IP Hamiltonian.
For a self-consistent theory, one would include electron-hole interaction effects through the self-energy.
Therefore, a correction term to the velocity operator might come from the effective single-particle Hamiltonian.
A fully self-consistent treatment is beyond this work and will be left in future work.

\section{Conclusion}
\label{sec:conclusion}
In summary, we derive the expression for second order optical responses including excitonic effects with a diagrammatic approach.
Our approach extends the previous derivation for the IP cases by dressing the electron-photon coupling vertex with the electron-hole ladder diagrams.
It is known that excitonic effects are strong for low dimensional materials. Hence, we expect that excitonic effects are essential to describe nonlinear optical responses. 
First-principle calculations of second order optical responses including excitonic effects can be straightforwardly performed by implementing the derived expression. All the ingredients can be obtained from standard density functional theory packages and software with BSE solvers implemented.
Although we only focus on the second order response, the diagrammatic rules provided here can be readily applied to higher order responses.

\section*{Acknowledgement}

Y.-H. C. and Y.T.C thank Hsin Lin for the discussion. Y.-H. C. thanks Jiawei Ruan and Prof. Steven G. Louie for collaborations on related projects. This work was supported by the National Science and Technology Council of Taiwan under grant no. 112-2112-M-001-048-MY3. We acknowledge the use of computational resources at the National Center for High-performance Computing (NCHC) in Taiwan.

\appendix
\onecolumngrid

\section{Green function and correlation function}
\label{sec:appendixA}

Throughout Appendix~\ref{sec:appendixA} we set $\hbar=1$ to simplify the bookkeeping and the Einstein summation convention is implied for band and momentum indices. Following Ref.~\cite{Parker2019}, the non-interacting single particle Green function is
\begin{align*}
    G_{b}(\omega)=\frac{1}{\omega-\epsilon_b}.
\end{align*}
It is frequency integral is
\[
I_{1}(\omega_{1})=\int\frac{d\omega}{2\pi}G_{a}(\omega)=f(\epsilon_a),
\]
where $f(\epsilon_a)$ is the Fermi-Dirac function.
The convolution of two and three Green function are
\[
I_{2}(\omega_{1})=\int\frac{d\omega}{2\pi}G_{a}(\omega)G_{b}(\omega+\omega_{1}),
\]
\[
I_{3}(\omega_{1},\omega_{2})=\int\frac{d\omega}{2\pi}G_{a}(\omega)G_{b}(\omega+\omega_{1})G_{c}(\omega+\omega_{1}+\omega_{2}),
\]
respectively.
The evaluation of these integrals can be done by working with Matsubara frequencies then perform analytical continuation back to the real frequncies~\cite{Mahan}.
For $I_{2}$, we consider
\[
S_{2}(i\omega_{1})=\frac{1}{\beta}\sum_{n}\frac{1}{z_{n}-\epsilon_{a}}\frac{1}{z_{n}+i\omega_{1}-\epsilon_{b}},
\]
where $z_{n}=i(2n+1)\pi/\beta$. The summation can be done by considering
a contour integral of a function
\[
0=\int\frac{dz}{2\pi i}f(z)F(z)
\]
with $f(z)=\frac{1}{e^{\beta z}+1}$ and 
\[
F(z)=\frac{1}{z-\epsilon_{a}}\frac{1}{z+i\omega_{1}-\epsilon_{b}}.
\]
The function $f(z)F(z)$ has poles and contribute to the contour integral
at:
\[
z=z_{n},\qquad R_{1}=-\frac{1}{\beta}F(z_{n})
\]
\[
z=\epsilon_{a},\qquad R_{2}=\frac{f(\epsilon_{a})}{\epsilon_{a}-\epsilon_{b}+i\omega_{1}}
\]
\[
z=\epsilon_{b}-i\omega_{1},\qquad R_{3}=\frac{f(\epsilon_{b})}{\epsilon_{b}-\epsilon_{a}-i\omega_{1}}.
\]
We note that the residue of Fermi function is $-1/\beta$ so we have
\begin{align*}
S_{2}(i\omega_{1}) & =\frac{1}{\beta}F(z_{n})=\frac{f(\epsilon_{a})}{\epsilon_{a}-\epsilon_{b}+i\omega_{1}}-\frac{f(\epsilon_{b})}{\epsilon_{a}-\epsilon_{b}+i\omega_{1}}\\
 & =-\frac{f_{ba}}{i\omega_{1}-\epsilon_{ba}},
\end{align*}
where we define $f_{ab}=f(\epsilon_a)-f(\epsilon_b)$ and $\epsilon_{ab}=\epsilon_a-\epsilon_b$.
For $I_{3}$, we consider the following function
\[
F_{3}(z)=\frac{1}{z-\epsilon_{a}}\frac{1}{z+i\omega_{1}-\epsilon_{b}}\frac{1}{z+i\omega_{1}+i\omega_{2}-\epsilon_{c}}.
\]
The contour integral has three pieces
\[
z=z_{n},\qquad R_{1}=-\frac{1}{\beta}F_{3}(z_{n})
\]
\[
z=\epsilon_{a},\qquad R_{2}=\frac{f(\epsilon_{a})}{(\epsilon_{ab}+i\omega_{1})(\epsilon_{ac}+i\omega_{12})}
\]
\[
z=\epsilon_{b}-i\omega_{1},\qquad R_{3}=\frac{f(\epsilon_{b})}{(\epsilon_{ba}-i\omega_{1})(\epsilon_{bc}+i\omega_{2})}
\]
\[
z=\epsilon_{c}-i\omega_{1}-i\omega_{2},\qquad R_{4}=\frac{f(\epsilon_{c})}{(\epsilon_{ca}-i\omega_{12})(\epsilon_{cb}-i\omega_{2})}.
\],
where $\omega_{12}=\omega_1+\omega_2$.
The sum of them gives
\begin{align}
I_{3}(i\omega_{1},i\omega_{2}) & =\frac{f(\epsilon_{a})}{(\epsilon_{ba}-i\omega_{1})(\epsilon_{ca}-i\omega_{12})}+\frac{f(\epsilon_{b})}{(\epsilon_{ba}-i\omega_{1})(\epsilon_{bc}+i\omega_{2})}+\frac{-f(\epsilon_{c})}{(\epsilon_{ca}-i\omega_{12})(\epsilon_{bc}+i\omega_{2})}\nonumber\\
 & =\frac{f(\epsilon_{a})(\epsilon_{bc}+i\omega_{2})+f(\epsilon_{b})(\epsilon_{ca}-i\omega_{12})-f(\epsilon_{c})(\epsilon_{ba}-i\omega_{1})}{(\epsilon_{ba}-i\omega_{1})(\epsilon_{bc}+i\omega_{2})(\epsilon_{ca}-i\omega_{12})}\nonumber\\
 &=\frac{f_{ab}(i\omega_{2}-\epsilon_{cb})+f_{cb}(i\omega_{1}-\epsilon_{ba})}{(i\omega_{1}-\epsilon_{ba})(i\omega_{2}-\epsilon_{cb})(i\omega_{12}-\epsilon_{ca})}.
 \label{eq:I3}
\end{align}

\subsection*{Two particle correlation function}

To compute the non-interacting two particle correlation function, we start from the Matsubara component,
\[
L_{ab}^{M}(i\omega_{p})=\frac{i}{\beta}\sum_{m}G_{b}^{M}(i\omega_{m})G_{a}^{M}(i\omega_{m}+i\omega_{p}).
\]
Using the result in the previous section, we have
\[
L_{ab}^{M}(i\omega_{p})=i\frac{f_{ba}}{i\omega_{p}-\epsilon_{ab}}.
\]
The retarded component can be obtained by an analytical continuation, which reads,
\[
L_{0,ab}^{R}(\omega)=i\frac{f_{ba}}{\omega+i\eta-\epsilon_{ab}},
\]
where the subscript $0$ indicates that it is the non-interacting correlation function.

Interacting electron-hole correlation function can be constructed from the BSE solutions~\cite{Rohlfing2000,Onida2002,Perfetto2016,Sven2020}. 
With Tamm-Dancoff approximation, we have
\[
H_{cvc'v'}Y^\lambda_{c'v'}=(\omega_{cv}-K_{cvc'v'})Y^\lambda_{c'v'}=\Omega_{\lambda}Y^\lambda_{c'v'} 
\]
\[
H_{vcv'c'}X^\lambda_{c'v'}=(\omega_{vc}+K_{vcv'c'})X^\lambda_{v'c'}=-\Omega_{\lambda}X^\lambda_{v'c'} 
\]
where $\Omega_\lambda$ is the eigenvalue, $Y^\lambda$ and $X^\lambda$ are eigenvectors for resonant and anti-resonant sectors, respectively.
We see that $H_{vcv'c'}=-H_{cvc'v'}^{*}$ and $X^\lambda_{v'c'}=Y_{c'v'}^{\lambda*}$.
The interacting correlation funciton can then be constructed as,
\[
L_{ij\mathbf{k}nm\mathbf{k}'}^{R}(\omega)=i\sum_{\lambda}\left[\bar{f}_{i}f_{j}\bar{f}_{n}f_{m}\frac{Y^\lambda_{ij\mathbf{k}}Y_{nm\mathbf{k}'}^{\lambda*}}{\omega-\Omega_{\lambda}+i\eta}-f_{i}\bar{f}_{j}\bar{f}_{m}f_{n}\frac{Y_{ji\mathbf{k}}^{\lambda*}Y^\lambda_{mn\mathbf{k}'}}{\omega+\Omega_{\lambda}+i\eta}\right],
\]
where we define $\bar{f}=1-f$.

\section{Derivation of the second order conductivity tensor}

In the first part of this section, we give the detailed derivations of the second order optical conductivity within IP approximation. The derivation including electron-hole interactions is given in the second part. These derivations demonstrate the Feynmann rules listed in Ref.~\cite{Parker2019} and in our work. 

\subsection*{IP}
\label{sec:appendixB1}
For the second order response, the conductivity tensor $\sigma^{\mu\alpha\beta}(\omega;\omega_{1},\omega_{2})$
can be computed from the second derivative of the current density with respect to the external fields, $\frac{\delta J^\mu(t)}{\delta E^\alpha(t_{1})\delta E^\beta(t_{2})}|_{E=0}$. From Eq.~\ref{eq:J} we can identify that there are two operators which consist of expansion of the external field. One is the velocity operator associated with the current operator and the other is the coupling Hamiltonian, $H_E$. In total there are four terms contributing to the second order conductivity, 
\begin{equation}
\frac{\delta^{2}v_{E}^{\mu}(t)}{\delta E^{\beta}(t_{2})\delta E^{\alpha}(t_{1})}-v_{E}^{\mu}(t)\frac{\delta^{2}\int dt'H_{E}(t')}{\delta E^{\beta}(t_{2})\delta E^{\alpha}(t_{1})}
-\frac{\delta v_{E}^{\mu}(t)}{\delta E^{\beta}(t_{2})}\frac{\delta\int dt'H_{E}(t')}{\delta E^{\alpha}(t_{1})}+\frac{1}{2!}v_{E}^{\mu}(t)\frac{\delta^{2}(\int dt''H_{E}(t''))^{2}}{\delta E^{\alpha}(t_{1})\delta E^{\beta}(t_{2})},
\label{eq:IP4}
\end{equation}
where $\mathbf{E}=0$ is set after taking the derivatives.
As we discussed in the main text, we identify the vertex associated with the current operator as the out-going vertex and those associated with the $H_E$ as the incoming vertex. The number of photon lines on a vertex is dertermined by the order of derivatives with respect to the external field.
Therefore the first term correspond to the diagram with three photon lines on the out-going vertex, the third term is the diagram with two photon lines on the out-going vertex and one photon line on the incoming vertex. The second and fourth term both have two incoming vertex and one out-going vertex. While the second term has two photon lines on the incoming vertex, the fourth term has one photon line on each incoming vertex. We note that the functional derivative can be taken in a different order, which corresponds to the symmetry of exchanging the two external fields.

We are interested in the second and the third term which generate diagrams with two incoming or two outgoing photons on the same vertex.
The third term can be obtained by combining terms we calculated for the first order conductivity.
Using $\frac{\delta}{\delta E(t)}=\int d\omega'\,e^{i\omega't}\frac{\delta}{\delta E(\omega')}$, we have
\begin{align*}
 & -\frac{\delta v_{E}^{\mu}(t)}{\delta E^{\beta}(t_{2})}\frac{\delta}{\delta E^{\alpha}(t_{1})}\int dt'H_{E}(t')\\
 & =-\int d\omega_{1}\,e^{i\omega_{1}t_{2}}\frac{\delta}{\delta E^{\beta}(\omega_{1})}\,\left[\sum_{n=0}^{\infty}\frac{1}{n!}\prod_{k=1}^{n}\int\frac{d\omega_{k}}{2\pi}\,e^{-i\omega_{k}t}\left(\frac{i\text{e}}{\hbar\omega_{k}}\right)E^{\alpha_{k}}(\omega_{k})\hat{h}^{\mu\alpha_{1}...\alpha_{n}}\right]\\
 & \times\int dt'\int d\omega_{2}\,e^{i\omega_{2}t_{1}}\frac{\delta}{\delta E^{\alpha}(\omega_{2})}\,\left[\sum_{n=1}^{\infty}\frac{1}{n!}\prod_{k=1}^{n}\int\frac{d\omega_{k}}{2\pi}\,e^{-i\omega_{k}t'}\left(\frac{i\text{e}}{\hbar\omega_{k}}\right)E^{\alpha_{k}}(\omega_{k})\hat{h}^{\alpha_{1}...\alpha_{n}}\right]\\
 & =-\int dt'\int\frac{d\omega_{1}}{2\pi}\int\frac{d\omega_{2}}{2\pi}\,e^{-i\omega_{2}(t'-t_{1})}\,e^{-i\omega_{1}(t-t_{2})}\left(\frac{i\text{e}}{\hbar\omega_{1}}\right)\hat{h}^{\mu\beta}(t)\left(\frac{i\text{e}}{\hbar\omega_{2}}\right)\hat{h}^{\alpha}(t')
\end{align*}
The term with the second derivative in the second term of Eq.~\ref{eq:IP4} reads, 
\begin{align*}
 & \frac{\delta^{2}}{\delta E^{\beta}(t_{2})\delta E^{\alpha}(t_{1})}\int dt'H_{E}(t')\\
= & \int d\omega''e^{i\omega''t_{2}}\frac{\delta}{\delta E^{\beta}(\omega'')}\int d\omega'e^{i\omega't_{1}}\frac{\delta}{\delta E^{\alpha}(\omega')}\,\\
 & \times\int dt'\sum_{n=1}^{\infty}\frac{1}{n!}\prod_{k=1}^{n}\int\frac{d\omega_{k}}{2\pi}\,e^{-i\omega_{k}t'}\left(\frac{i\text{e}}{\hbar\omega_{k}}\right)E^{\alpha_{k}}(\omega_{k})\hat{h}^{\alpha_{1}...\alpha_{n}}\\
= & \frac{1}{2!}\int d\omega''e^{i\omega''t_{2}}\frac{\delta}{\delta E^{\beta}(\omega'')}\int d\omega'e^{i\omega't_{1}}\frac{\delta}{\delta E^{\alpha}(\omega')}\int dt'\int\frac{d\omega_{1}}{2\pi}\int\frac{d\omega_{2}}{2\pi}\\
 & \times e^{-i\omega_{1}t'}e^{-i\omega_{2}t'}\left(\frac{i\text{e}}{\hbar\omega_{1}}\right)\left(\frac{i\text{e}}{\hbar\omega_{2}}\right)E^{\alpha_{1}}(\omega_{1})E^{\alpha_{2}}(\omega_{2})\hat{h}^{\alpha_{1}\alpha_{2}}\\
= & \frac{1}{2!}\int dt'\int\frac{d\omega''}{2\pi}e^{i\omega''(t_{2}-t')}\int\frac{d\omega'}{2\pi}\,e^{i\omega'(t_{1}-t')}\left(\frac{i\text{e}}{\hbar\omega''}\right)\left(\frac{i\text{e}}{\hbar\omega'}\right)\left(\hat{h}^{\alpha\beta}(t')+\hat{h}^{\beta\alpha}(t')\right).
\end{align*}
Using $v_{E}^{\mu}=h^{\mu}$ at the zeroth order, the second term is,
\[
-h^{\mu}(t)\frac{1}{2!}\int dt'\int\frac{d\omega''}{2\pi}e^{i\omega''(t_{2}-t')}\int\frac{d\omega'}{2\pi}\,e^{i\omega'(t_{1}-t')}\left(\frac{i\text{e}}{\hbar\omega''}\right)\left(\frac{i\text{e}}{\hbar\omega'}\right)\left(\hat{h}^{\alpha\beta}(t')+\hat{h}^{\beta\alpha}(t')\right).
\]

The Fourier transformation of the expectation value of the third term is
\begin{align*}
 & \sigma_{IP}^{\mu\alpha\beta,3}(\omega;\omega_{1},\omega_{2})\frac{\delta(\omega-\omega_{1}-\omega_{2})}{2\pi}\\
 & =-\text{e}\int dte^{i\omega t}\int\frac{dt_{1}}{2\pi}e^{-i\omega_{1}t_{1}}\int\frac{dt_{2}}{2\pi}e^{-i\omega_{2}t_{2}}\\
 & \times\int dt'\int\frac{d\omega_{3}}{2\pi}\int\frac{d\omega_{4}}{2\pi}\,e^{-i\omega_{4}(t'-t_{1})}\,e^{-i\omega_{3}(t-t_{2})}\left(\frac{i\text{e}}{\hbar\omega_{3}}\right)\left(\frac{i\text{e}}{\hbar\omega_{4}}\right)\expval{\hat{h}^{\mu\beta}(t)\hat{h}^{\alpha}(t')}\\
 & =-\text{e}\int dt\int\frac{dt_{1}}{2\pi}e^{i(-\omega_{1}+\omega_{4})t_{1}}\int\frac{dt_{2}}{2\pi}e^{i(-\omega_{2}+\omega_{3})t_{2}}\\
 & \times\int dt'\int\frac{d\omega_{3}}{2\pi}\int\frac{d\omega_{4}}{2\pi}\,e^{-i\omega_{4}t'}\,e^{-i(\omega_{3}-\omega)t}\left(\frac{i\text{e}}{\hbar\omega_{3}}\right)\left(\frac{i\text{e}}{\hbar\omega_{4}}\right)\expval{\hat{h}^{\mu\beta}(t)\hat{h}^{\alpha}(t')}\\
 & =\frac{1}{(2\pi)^{2}}\frac{\text{e}^{3}}{\hbar^{2}\omega_{2}\omega_{1}}\int dt\int dt'\,e^{i\omega_{1}t'}\,e^{-i(\omega_{2}-\omega)t}h_{ab}^{\mu\beta}h_{cd}^{\alpha}\left\langle c_{a}^{\dagger}(t)c_{b}(t)c_{c}^{\dagger}(t')c_{d}(t')\right\rangle \\
 & =-\frac{1}{(2\pi)^{2}}\frac{\text{e}^{3}}{\hbar^{2}\omega_{2}\omega_{1}}\int dt\int dt'\,e^{i\omega_{1}t'}\,e^{-i(\omega_{2}-\omega)t}h_{ab}^{\mu\beta}h_{ba}^{\alpha}G_{b}(t,t')G_{a}(t',t)\\
 & =-\frac{1}{(2\pi)^{2}}\frac{\text{e}^{3}}{\hbar^{2}\omega_{2}\omega_{1}}\int dt\int dt'\,e^{i\omega_{1}t'}\,e^{-i(\omega_{2}-\omega)t}h_{ab}^{\mu\beta}h_{ba}^{\alpha}\int\frac{d\omega'}{2\pi}G_{b}(\omega')e^{-i\omega'(t-t')}\int\frac{d\omega''}{2\pi}e^{-i\omega''(t'-t)}G_{a}(\omega'')\\
 & =-\frac{1}{(2\pi)^{4}}\frac{\text{e}^{3}}{\hbar^{2}\omega_{2}\omega_{1}}\int d\omega'\int d\omega''\int dt\int dt'\,e^{i(\omega_{1}+\omega'-\omega'')t'}\,e^{-i(\omega_{2}-\omega+\omega'-\omega'')t}h_{ab}^{\mu\beta}h_{ba}^{\alpha}G_{b}(\omega')G_{a}(\omega'')\\
 & =-\frac{\text{e}^{3}}{\hbar^{2}\omega_{2}\omega_{1}}\frac{\delta(\omega-\omega_{1}-\omega_{2})}{2\pi}\int\frac{d\omega'}{2\pi}h_{ab}^{\mu\beta}h_{ba}^{\alpha}G_{b}(\omega')G_{a}(\omega_{1}+\omega'),
\end{align*}
where the superscript 3 indicates that it is the third term in Eq.~\ref{eq:IP4} and in the sixth line only the connected diagram is considered. 
For the second term, we have
\begin{align*}
 & \sigma_{IP}^{\mu\alpha\beta,2}(\omega;\omega_{1},\omega_{2})\frac{\delta(\omega-\omega_{1}-\omega_{2})}{2\pi}\\
 & =-\frac{\text{e}}{2!}\int dte^{i\omega t}\int\frac{dt_{1}}{2\pi}\int\frac{dt_{2}}{2\pi}\int dt'\int\frac{d\omega''}{2\pi}\int\frac{d\omega'}{2\pi}e^{i(-\omega_{2}+\omega'')t_{2}}e^{i(-\omega_{1}+\omega')t_{1}}e^{-i(\omega'+\omega'')t'}\,\\
 & \times\left(\frac{i\text{e}}{\hbar\omega''}\right)\left(\frac{i\text{e}}{\hbar\omega'}\right)\expval{\hat{h}^{\mu}(t)\hat{h}^{\alpha\beta}(t')}\\
 & =\frac{\text{e}^{3}}{\hbar^{2}\omega_{1}\omega_{2}}\frac{1}{2!(2\pi)^{2}}\int dte^{i\omega t}\int dt'e^{-i(\omega_{1}+\omega_{2})t'}\,\expval{\hat{h}^{\mu}(t)\hat{h}^{\alpha\beta}(t')}\\
 & =\frac{1}{2!(2\pi)^{2}}\frac{\text{e}^{3}}{\hbar^{2}\omega_{1}\omega_{2}}\int dte^{i\omega t}\int dt'e^{-i(\omega_{1}+\omega_{2})t'}\,h_{ab}^{\mu}h_{cd}^{\alpha\beta}\left\langle c_{a}^{\dagger}(t)c_{b}(t)c_{c}^{\dagger}(t')c_{d}(t')\right\rangle \\
 & =-\frac{1}{2!(2\pi)^{2}}\frac{\text{e}^{3}}{\hbar^{2}\omega_{1}\omega_{2}}\int dte^{i\omega t}\int dt'e^{-i(\omega_{1}+\omega_{2})t'}\,h_{ab}^{\mu}h_{cd}^{\alpha\beta}G_{bc}(t,t')G_{da}(t',t)\\
 & =-\frac{1}{2!(2\pi)^{2}}\frac{\text{e}^{3}}{\hbar^{2}\omega_{1}\omega_{2}}\int dte^{i\omega t}\int dt'e^{-i(\omega_{1}+\omega_{2})t'}\,h_{ab}^{\mu}h_{ba}^{\alpha\beta}\int\frac{d\omega'}{2\pi}G_{b}(\omega')e^{-i\omega'(t-t')}\int\frac{d\omega''}{2\pi}e^{-i\omega''(t'-t)}G_{a}(\omega'')\\
 & =-\frac{1}{2!(2\pi)^{2}}\frac{\text{e}^{3}}{\hbar^{2}\omega_{1}\omega_{2}}\int dt\int dt'e^{i(-\omega_{1}-\omega_{2}+\omega'-\omega'')t'}\,h_{ab}^{\mu}h_{ba}^{\alpha\beta}\int\frac{d\omega'}{2\pi}\int\frac{d\omega''}{2\pi}e^{i(\omega''+\omega-\omega')t}G_{b}(\omega')G_{a}(\omega'')\\
 & =-\frac{\text{e}^{3}}{2\hbar^{2}\omega_{1}\omega_{2}}\frac{\delta(\omega-\omega_{1}-\omega_{2})}{2\pi}h_{ab}^{\mu}h_{ba}^{\alpha\beta}\int\frac{d\omega''}{2\pi}G_{b}(\omega''+\omega_{1}+\omega_{2})G_{a}(\omega'').
\end{align*}
These results agree with Eq. (41) in Ref.~\cite{Parker2019}

\subsection*{With electron-hole interactions}

For the second order response, there are four diagrams and their permutations.
Vertex corrections of the first three diagrams in Ref.~\cite{Parker2019} are similar to those in the linear response. 
The second term in Eq. (41) in Ref.~\cite{Parker2019} gets vertex correction on the out-going vertex with two photon lines,
\begin{align}
 & \sigma_{eh}^{\mu\alpha\beta,2}\nonumber\\
 & =\frac{-\text{e}^{3}}{\hbar^{2}\omega_{1}\omega_{2}}h_{ab}^{\alpha}\int d\omega'G_{b}(\omega')G_{a}(\omega'+\omega_{1})\Tilde{h}_{ab}^{\mu\beta}(\omega_{1})\nonumber\\
 & =\frac{-\text{e}^{3}}{\hbar^{2}\omega_{1}\omega_{2}}h_{ab}^{\alpha}L_{0,ab}(\omega_{1})h_{cd}^{\mu\beta*}L_{cd,ab}(\omega_{1})L_{0,ab}^{-1}(\omega_{1})\nonumber\\
 & =\frac{-\text{e}^{3}}{\hbar^{2}\omega_{1}\omega_{2}}h_{ab}^{\alpha}L_{cd,ab}(\omega_{1})h_{cd}^{\mu\beta*}\nonumber\\
 & =\frac{-\text{e}^{3}}{\hbar^{2}\omega_{1}\omega_{2}}\sum_{\lambda}h_{ab}^{\alpha}\left[\bar{f}_{c}f_{d}\bar{f}_{a}f_{b}\frac{Y_{cd\mathbf{k}}^{\lambda}Y_{ab\mathbf{k}'}^{\lambda*}}{\hbar\omega_{1}-\Omega_{\lambda}+i\eta}-f_{c}\bar{f}_{d}\bar{f}_{a}f_{b}\frac{Y_{dc\mathbf{k}}^{\lambda*}Y_{ba\mathbf{k}'}^{\lambda}}{\hbar\omega_{1}+\Omega_{\lambda}+i\eta}\right]h_{cd}^{\mu\beta*}\nonumber\\
 & =\frac{-\text{e}^{3}}{\hbar^{2}\omega_{1}\omega_{2}}\sum_{\lambda}\left[\frac{d_{\lambda}^{\alpha}d_{\lambda}^{\mu\beta*}}{\hbar\omega_{1}-\Omega_{\lambda}+i\eta}-\frac{d_{\lambda}^{\alpha*}d_{\lambda}^{\mu\beta}}{\hbar\omega_{1}+\Omega_{\lambda}+i\eta}\right],
 \label{eq:eh22}
\end{align}
where we denote dressed operators with a tilde and the superscript 2 indicates that it is the second term in Eq. (41) in Ref.~\cite{Parker2019}; In the third line we use the inverse of the bare two particle correlation function,
\[
L_{0,ab}^{-1}=\frac{\hbar\omega-\epsilon_{ab}}{f_{ba}},
\]
and in the last line we define $d_{\lambda}^{\alpha\beta}=h_{ab}^{\alpha\beta}Y_{ab\mathbf{k}}^{\lambda*}$.
We can also dress the incoming vertex,
\begin{align*}
 & \sigma_{eh}^{\mu\alpha\beta,2}\\
 & =\frac{-\text{e}^{3}}{\hbar^{2}\omega_{1}\omega_{2}}\Tilde{h}_{ab}^{\alpha}(\omega_{1})\int d\omega'G_{b}(\omega')G_{a}(\omega'+\omega_{1})h_{ba}^{\mu\beta}\\
 & =\frac{-\text{e}^{3}}{\hbar^{2}\omega_{1}\omega_{2}}L_{0,ab}^{-1}(\omega_{1})L_{ab,cd}(\omega_{1})h_{cd}^{\alpha}L_{0,ab}(\omega_{1})h_{ba}^{\mu\beta}\\
 & =\frac{-\text{e}^{3}}{\hbar^{2}\omega_{1}\omega_{2}}h_{cd}^{\alpha}L_{ab,cd}(\omega_{1})h_{ba}^{\mu\beta},
\end{align*}
which is equivalent to Eq.~\ref{eq:eh22}. 

For the third diagram in Eq.~(41) in Ref.~\cite{Parker2019}, we dress the incoming vertex and get 
\begin{align*}
 & \sigma_{eh}^{\mu\alpha\beta,3}\\
 & =\frac{-\text{e}^{3}}{2\hbar^{2}\omega_{1}\omega_{2}}\Tilde{h}_{ab}^{\alpha\beta}(\omega)\int d\omega'G_{b}(\omega')G_{a}(\omega'+\omega_{12})h_{ba}^{\mu}\\
 & =\frac{-\text{e}^{3}}{2\hbar^{2}\omega_{1}\omega_{2}}L_{0,ab}^{-1}(\omega_{12})L_{ab,cd}(\omega_{12})h_{cd}^{\alpha\beta}L_{0,ab}(\omega_{12})h_{ba}^{\mu}\\
 & =\frac{-\text{e}^{3}}{2\hbar^{2}\omega_{1}\omega_{2}}h_{cd}^{\alpha\beta}L_{ab,cd}(\omega_{12})h_{ba}^{\mu}\\
 & =\frac{-\text{e}^{3}}{2\hbar^{2}\omega_{1}\omega_{2}}\sum_{s}\left[h_{cd}^{\alpha\beta}\frac{Y_{ab\mathbf{k}}^{\lambda}Y_{cd\mathbf{k}'}^{\lambda*}}{\hbar\omega-\Omega_{\lambda}+i\eta}h_{ba}^{\mu}-h_{cd}^{\alpha\beta}\frac{Y_{ba\mathbf{k}}^{\lambda*}Y_{dc\mathbf{k}'}^{\lambda}}{\hbar\omega+\Omega_{\lambda}+i\eta}h_{ba}^{\mu}\right]\\
 & =\frac{-\text{e}^{3}}{2\hbar^{2}\omega_{1}\omega_{2}}\sum_{s}\left[\frac{d_{\lambda}^{\alpha\beta}d_{\lambda}^{\mu*}}{\hbar\omega-\Omega_{\lambda}+i\eta}-\frac{d_{\lambda}^{\alpha\beta*}d_{\lambda}^{\mu}}{\hbar\omega+\Omega_{\lambda}+i\eta}\right].
\end{align*}

The new diagram at the second order response is the triangle diagram with three
vertice. As discussed in the main text, we approximate the interacting three particle correlation function with the e-h ladder diagram and obtain three diagrams, each of which has two dressed vertice and one bare vertex. 
For the first derived diagram we dress one of the incoming vertice and the outgoing vertex,
\begin{align*}
 & \sigma_{eh}^{\mu\alpha\beta,4.1}\\
 & =\frac{-\text{e}^{3}}{\hbar^{2}\omega_{1}\omega_{2}}\Tilde{h}_{ba}^{\alpha}(\omega_{1})h_{cb}^{\beta}\Tilde{h}_{ca}^{\mu}(\omega)I_{abc}(\omega_{1},\omega_{2})\\
 & =\frac{-\text{e}^{3}}{\hbar^{2}\omega_{1}\omega_{2}}L_{0,ba}^{-1}(\omega_{1})L_{ba,ij}(\omega_{1})h_{ij}^{\alpha}h_{cb}^{\beta}h_{ns}^{\mu*}L_{ns,ca}(\omega)L_{0,ca}^{-1}(\omega)I_{abc}(\omega_{1},\omega_{2})\\
 & =\frac{-\text{e}^{3}}{\hbar^{2}\omega_{1}\omega_{2}}\frac{\omega_{1}-\epsilon_{ba}}{f_{ab}}L_{ba,ij}(\omega_{1})h_{ij}^{\alpha}h_{cb}^{\beta}h_{ns}^{\mu*}L_{ns,ca}(\omega)\frac{\omega-\epsilon_{ca}}{f_{ac}}\frac{f_{ab}(\omega_{2}-\epsilon_{cb})+f_{cb}(\omega_{1}-\epsilon_{ba})}{(\omega_{1}-\epsilon_{ba})(\omega-\epsilon_{ca})(\omega_{2}-\epsilon_{cb})}\\
 & =\frac{-\text{e}^{3}}{\hbar^{2}\omega_{1}\omega_{2}}\sum_{s\lambda}h_{ij}^{\alpha}h_{cb}^{\beta}h_{ns}^{\mu*}\frac{1}{f_{ab}}\frac{1}{f_{ac}}\frac{f_{ab}(\omega_{2}-\epsilon_{cb})+f_{cb}(\omega_{1}-\epsilon_{ba})}{(\omega_{2}-\epsilon_{cb})}\\
 & \times\left(\frac{Y_{ba\mathbf{k}}^{s}Y_{ij\mathbf{k}'}^{s*}}{\hbar\omega_{1}-\Omega_{s}+i\eta}\frac{Y_{ns\mathbf{k}}^{\lambda}Y_{ca\mathbf{k}'}^{\lambda*}}{\hbar\omega-\Omega_{\lambda}+i\eta}+\frac{Y_{ab\mathbf{k}}^{s*}Y_{ji\mathbf{k}'}^{s}}{\hbar\omega_{1}+\Omega_{s}+i\eta}\frac{Y_{sn\mathbf{k}}^{\lambda*}Y_{ac\mathbf{k}'}^{\lambda}}{\hbar\omega+\Omega_{\lambda}+i\eta}\right)\\
 & =\frac{\text{e}^{3}}{\hbar^{2}\omega_{1}\omega_{2}}\sum_{s\lambda}h_{ij}^{\alpha}h_{cb}^{\beta}h_{ns}^{\mu*}\left(-\frac{Y_{ba\mathbf{k}}^{s}Y_{ij\mathbf{k}'}^{s*}}{\hbar\omega_{1}-\Omega_{s}+i\eta}\frac{Y_{ns\mathbf{k}}^{\lambda}Y_{ca\mathbf{k}'}^{\lambda*}}{\hbar\omega-\Omega_{\lambda}+i\eta}+\frac{Y_{ab\mathbf{k}}^{s*}Y_{ji\mathbf{k}'}^{s}}{\hbar\omega_{1}+\Omega_{s}+i\eta}\frac{Y_{sn\mathbf{k}}^{\lambda*}Y_{ac\mathbf{k}'}^{\lambda}}{\hbar\omega+\Omega_{\lambda}+i\eta}\right)\\
 & =\frac{\text{e}^{3}}{\hbar^{2}\omega_{1}\omega_{2}}\sum_{s\lambda}h_{cb}^{\beta}\left(-\frac{Y_{ba\mathbf{k}}^{s}d_{s}^{\alpha}}{\hbar\omega_{1}-\Omega_{s}+i\eta}\frac{d_{\lambda}^{\mu*}Y_{ca\mathbf{k}'}^{\lambda*}}{\hbar\omega-\Omega_{\lambda}+i\eta}+\frac{Y_{ab\mathbf{k}}^{s*}d_{s}^{\alpha*}}{\hbar\omega_{1}+\Omega_{s}+i\eta}\frac{d_{\lambda}^{\mu}Y_{ac\mathbf{k}'}^{\lambda}}{\hbar\omega+\Omega_{\lambda}+i\eta}\right),
\end{align*}
where the superscript 4.1 indicates that it is the first diagram derived from the triangle diagram, in the third line, we replace the bare three particle correlation function, $I_{abc}$ with Eq.~\ref{eq:I3} introduced in Appendix.~\ref{sec:appendixA}, in the sixth line we set the occupations to their equilibrium values, and in the fifth line, we replace the product of the two correlation functions with the following,
\begin{align*}
L_{ba,ij}(\omega_{1})L_{ns,ca}(\omega)
= \sum_{s\lambda}\left(\frac{Y_{ba\mathbf{k}}^{s}Y_{ij\mathbf{k}'}^{s*}}{\hbar\omega_{1}-\Omega_{s}+i\eta}\frac{Y_{ns\mathbf{k}}^{\lambda}Y_{ca\mathbf{k}'}^{\lambda*}}{\hbar\omega-\Omega_{\lambda}+i\eta}+\frac{Y_{ab\mathbf{k}}^{s*}Y_{ji\mathbf{k}'}^{s}}{\hbar\omega_{1}+\Omega_{s}+i\eta}\frac{Y_{sn\mathbf{k}}^{\lambda*}Y_{ac\mathbf{k}'}^{\lambda}}{\hbar\omega+\Omega_{\lambda}+i\eta}\right).
\end{align*}

For the second derived diagram, we dressed the other incoming vertex and the outgoing vertex, 
\begin{align*}
 & \sigma_{eh}^{\mu\alpha\beta,4.2}\\
 & =\frac{-\text{e}^{3}}{\hbar^{2}\omega_{1}\omega_{2}}h_{ba}^{\alpha}\Tilde{h}_{cb}^{\beta}(\omega_{2})\Tilde{h}_{ca}^{\mu}(\omega)I_{abc}(\omega_{1},\omega_{2})\\
 & =\frac{-\text{e}^{3}}{\hbar^{2}\omega_{1}\omega_{2}}h_{ba}^{\alpha}L_{0,cb}^{-1}(\omega_{2})L_{cb,lm}(\omega_{2})h_{lm}^{\beta}h_{ns}^{\mu*}L_{ns,ca}(\omega)L_{0,ca}^{-1}(\omega)I_{abc}(\omega_{1},\omega_{2})\\
 & =\frac{-\text{e}^{3}}{\hbar^{2}\omega_{1}\omega_{2}}\sum_{s\lambda}h_{ba}^{\alpha}\frac{\omega_{2}-\epsilon_{cb}}{f_{bc}}h_{lm}^{\beta}h_{ns}^{\mu*}\frac{\omega-\epsilon_{ca}}{f_{ac}}\frac{f_{ab}(\omega_{2}-\epsilon_{cb})+f_{cb}(\omega_{1}-\epsilon_{ba})}{(\omega_{1}-\epsilon_{ba})(\omega-\epsilon_{ca})(\omega_{2}-\epsilon_{cb})}L_{cb,lm}(\omega_{2})L_{ns,ca}(\omega)\\
 & =\frac{-\text{e}^{3}}{\hbar^{2}\omega_{1}\omega_{2}}\sum_{s\lambda}h_{ba}^{\alpha}h_{lm}^{\beta}h_{ns}^{\mu*}\frac{1}{f_{bc}}\frac{1}{f_{ac}}\frac{f_{ab}(\omega_{2}-\epsilon_{cb})+f_{cb}(\omega_{1}-\epsilon_{ba})}{(\omega_{1}-\epsilon_{ba})}\\
 & \times\left(\frac{Y_{cb\mathbf{k}}^{\lambda}Y_{lm\mathbf{k}'}^{\lambda*}}{\hbar\omega_{2}-\Omega_{\lambda}+i\eta}\frac{Y_{ns\mathbf{k}}^{s}Y_{ca\mathbf{k}'}^{s*}}{\hbar\omega-\Omega_{s}+i\eta}+\frac{Y_{bc\mathbf{k}}^{\lambda*}Y_{ml\mathbf{k}'}^{\lambda}}{\hbar\omega_{2}+\Omega_{\lambda}+i\eta}\frac{Y_{sn\mathbf{k}}^{s*}Y_{ac\mathbf{k}'}^{s}}{\hbar\omega+\Omega_{s}+i\eta}\right)\\
 & =\frac{-\text{e}^{3}}{\hbar^{2}\omega_{1}\omega_{2}}\sum_{s\lambda}h_{ba}^{\alpha}h_{lm}^{\beta}h_{ns}^{\mu*}\left(-\frac{Y_{cb\mathbf{k}}^{\lambda}Y_{lm\mathbf{k}'}^{\lambda*}}{\hbar\omega_{2}-\Omega_{\lambda}+i\eta}\frac{Y_{ns\mathbf{k}}^{s}Y_{ca\mathbf{k}'}^{s*}}{\hbar\omega-\Omega_{s}+i\eta}+\frac{Y_{bc\mathbf{k}}^{\lambda*}Y_{ml\mathbf{k}'}^{\lambda}}{\hbar\omega_{2}+\Omega_{\lambda}+i\eta}\frac{Y_{sn\mathbf{k}}^{s*}Y_{ac\mathbf{k}'}^{s}}{\hbar\omega+\Omega_{s}+i\eta}\right)\\
 & =\frac{-\text{e}^{3}}{\hbar^{2}\omega_{1}\omega_{2}}\sum_{s\lambda}h_{ba}^{\alpha}\left(-\frac{Y_{cb\mathbf{k}}^{\lambda}d_{\lambda}^{\beta}}{\hbar\omega_{2}-\Omega_{\lambda}+i\eta}\frac{d_{s}^{\mu*}Y_{ca\mathbf{k}'}^{s*}}{\hbar\omega-\Omega_{s}+i\eta}+\frac{Y_{bc\mathbf{k}}^{\lambda*}d_{\lambda}^{\beta*}}{\hbar\omega_{2}+\Omega_{\lambda}+i\eta}\frac{d_{s}^{\mu}Y_{ac\mathbf{k}'}^{s}}{\hbar\omega+\Omega_{s}+i\eta}\right).
\end{align*}
In the above, we replace the product of the correlation functions with,
\begin{equation*}
L_{cb,lm}(\omega_{2})L_{ns,ca}(\omega)=\sum_{s\lambda}\left(\frac{Y_{cb\mathbf{k}}^{\lambda}Y_{lm\mathbf{k}'}^{\lambda*}}{\hbar\omega_{2}-\Omega_{\lambda}+i\eta}\frac{Y_{ns\mathbf{k}}^{s}Y_{ca\mathbf{k}'}^{s*}}{\hbar\omega-\Omega_{s}+i\eta}+\frac{Y_{bc\mathbf{k}}^{\lambda*}Y_{ml\mathbf{k}'}^{\lambda}}{\hbar\omega_{2}+\Omega_{\lambda}+i\eta}\frac{Y_{sn\mathbf{k}}^{s*}Y_{ac\mathbf{k}'}^{s}}{\hbar\omega+\Omega_{s}+i\eta}\right).
\end{equation*}

For the third derived diagram we dress both incoming vertices, 
\begin{align*}
 & \sigma_{eh}^{\mu\alpha\beta,4.3}\\
 & =\frac{-\text{e}^{3}}{\hbar^{2}\omega_{1}\omega_{2}}\Tilde{h}_{ba}^{\alpha}(\omega_{1})\Tilde{h}_{cb}^{\beta}(\omega_{2})h_{ca}^{\mu}I_{abc}(\omega_{1},\omega_{2})\\
 & =\frac{-\text{e}^{3}}{\hbar^{2}\omega_{1}\omega_{2}}L_{0,ba}^{-1}(\omega_{1})L_{ba,ij}(\omega_{1})h_{ij}^{\alpha}L_{0,cb}^{-1}(\omega_{2})L_{cb,lm}(\omega_{2})h_{lm}^{\beta}h_{ac}^{\mu}I_{abc}(\omega_{1},\omega_{2})\\
 & =\frac{-\text{e}^{3}}{\hbar^{2}\omega_{1}\omega_{2}}\frac{\omega_{1}-\epsilon_{ba}}{f_{ab}}L_{ba,ij}(\omega_{1})h_{ij}^{\alpha}h_{lm}^{\beta}L_{cb,lm}(\omega_{2})\frac{\omega_{2}-\epsilon_{cb}}{f_{bc}}h_{ac}^{\mu}\frac{f_{ab}(\omega_{2}-\epsilon_{cb})+f_{cb}(\omega_{1}-\epsilon_{ba})}{(\omega_{1}-\epsilon_{ba})(\omega-\epsilon_{ca})(\omega_{2}-\epsilon_{cb})}\\
 & =\frac{-\text{e}^{3}}{\hbar^{2}\omega_{1}\omega_{2}} h_{ca}^{\mu*}h_{ij}^{\alpha}h_{lm}^{\beta}\frac{f_{ab}(\omega_{2}-\epsilon_{cb})+f_{cb}(\omega_{1}-\epsilon_{ba})}{(\omega-\epsilon_{ca})f_{ab}f_{bc}}\\
 & \times\sum_{s\lambda}\left(-\frac{Y_{ba\mathbf{k}}^{s}Y_{ij\mathbf{k}'}^{s*}}{\hbar\omega_{1}-\Omega_{s}+i\eta}\frac{Y_{bc\mathbf{k}}^{\lambda*}Y_{ml\mathbf{k}'}^{\lambda}}{\hbar\omega_{2}+\Omega_{\lambda}+i\eta}-\frac{Y_{ab\mathbf{k}}^{s*}Y_{ji\mathbf{k}'}^{s}}{\hbar\omega_{1}+\Omega_{s}+i\eta}\frac{Y_{cb\mathbf{k}}^{\lambda}Y_{lm\mathbf{k}'}^{\lambda*}}{\hbar\omega_{2}-\Omega_{\lambda}+i\eta}\right)\\
 & =\frac{-\text{e}^{3}}{\hbar^{2}\omega_{1}\omega_{2}} h_{ca}^{\mu*}h_{ij}^{\alpha}h_{lm}^{\beta}\sum_{s\lambda}\left(\frac{Y_{ba\mathbf{k}}^{s}Y_{ij\mathbf{k}'}^{s*}}{\hbar\omega_{1}-\Omega_{s}+i\eta}\frac{Y_{bc\mathbf{k}}^{\lambda*}Y_{ml\mathbf{k}'}^{\lambda}}{\hbar\omega_{2}+\Omega_{\lambda}+i\eta}-\frac{Y_{ab\mathbf{k}}^{s*}Y_{ji\mathbf{k}'}^{s}}{\hbar\omega_{1}+\Omega_{s}+i\eta}\frac{Y_{cb\mathbf{k}}^{\lambda}Y_{lm\mathbf{k}'}^{\lambda*}}{\hbar\omega_{2}-\Omega_{\lambda}+i\eta}\right)\\
 & =-\frac{\text{e}^{3}}{\hbar^{2}\omega_{1}\omega_{2}}\sum_{s\lambda}\left[\frac{h_{ca}^{\mu*}Y_{ba\mathbf{k}}^{s}Y_{bc\mathbf{k}}^{\lambda*}d_{s}^{\alpha}d_{\lambda}^{\beta*}}{(\hbar\omega_{1}-\Omega_{s}+i\eta)(\hbar\omega_{2}+\Omega_{\lambda}+i\eta)}-\frac{h_{ca}^{\mu*}Y_{ab\mathbf{k}}^{s*}Y_{cb\mathbf{k}}^{\lambda}d_{s}^{\alpha*}d_{\lambda}^{\beta}}{(\hbar\omega_{1}+\Omega_{s}+i\eta)(\hbar\omega_{2}-\Omega_{\lambda}+i\eta)}\right].
\end{align*}
In the fourth line we replace the product of two correlation functions using
\begin{align*}
 L_{ba,ij}(\omega_{1})L_{cb,lm}(\omega_{2})
= \sum_{s\lambda}\left(-\frac{Y_{ba\mathbf{k}}^{s}Y_{ij\mathbf{k}'}^{s*}}{\hbar\omega_{1}-\Omega_{s}+i\eta}\frac{Y_{bc\mathbf{k}}^{\lambda*}Y_{ml\mathbf{k}'}^{\lambda}}{\hbar\omega_{2}+\Omega_{\lambda}+i\eta}-\frac{Y_{ab\mathbf{k}}^{s*}Y_{ji\mathbf{k}'}^{s}}{\hbar\omega_{1}+\Omega_{s}+i\eta}\frac{Y_{cb\mathbf{k}}^{\lambda}Y_{lm\mathbf{k}'}^{\lambda*}}{\hbar\omega_{2}-\Omega_{\lambda}+i\eta}\right).
\end{align*}
Combining all three derived diagrams and explicitly symmetrizing them by adding terms with exchanged indices and frequencies, $\alpha\leftrightarrow\beta$, $\omega_{1}\leftrightarrow\omega_{2}$, we get
\begin{align*}
 & \sigma_{eh}^{\mu\alpha\beta,4}\\
 & =\frac{\text{e}^{3}}{\hbar^{2}\omega_{1}\omega_{2}}\sum_{s\lambda}\left(-\frac{h_{cb}^{\beta}Y_{ba\mathbf{k}}^{s}d_{s}^{\alpha}}{\hbar\omega_{1}-\Omega_{s}+i\eta}\frac{d_{\lambda}^{\mu*}Y_{ca\mathbf{k}'}^{\lambda*}}{\hbar\omega-\Omega_{\lambda}+i\eta}+\frac{h_{cb}^{\beta}Y_{ab\mathbf{k}}^{s*}d_{s}^{\alpha*}}{\hbar\omega_{1}+\Omega_{s}+i\eta}\frac{d_{\lambda}^{\mu}Y_{ac\mathbf{k}'}^{\lambda}}{\hbar\omega+\Omega_{\lambda}+i\eta}\right)\\
 & +\frac{\text{e}^{3}}{\hbar^{2}\omega_{1}\omega_{2}}\sum_{s\lambda}\left(-\frac{h_{cb}^{\alpha}Y_{ba\mathbf{k}}^{s}d_{s}^{\beta}}{\hbar\omega_{2}-\Omega_{s}+i\eta}\frac{d_{\lambda}^{\mu*}Y_{ca\mathbf{k}'}^{\lambda*}}{\hbar\omega-\Omega_{\lambda}+i\eta}+\frac{h_{cb}^{\alpha}Y_{ab\mathbf{k}}^{s*}d_{s}^{\beta*}}{\hbar\omega_{2}+\Omega_{s}+i\eta}\frac{d_{\lambda}^{\mu}Y_{ac\mathbf{k}'}^{\lambda}}{\hbar\omega+\Omega_{\lambda}+i\eta}\right)\\
 & -\frac{\text{e}^{3}}{\hbar^{2}\omega_{1}\omega_{2}}\sum_{s\lambda}\left(-\frac{h_{ba}^{\alpha}Y_{cb\mathbf{k}}^{\lambda}d_{\lambda}^{\beta}}{\hbar\omega_{2}-\Omega_{\lambda}+i\eta}\frac{d_{s}^{\mu*}Y_{ca\mathbf{k}'}^{s*}}{\hbar\omega-\Omega_{s}+i\eta}+\frac{h_{ba}^{\alpha}Y_{bc\mathbf{k}}^{\lambda*}d_{\lambda}^{\beta*}}{\hbar\omega_{2}+\Omega_{\lambda}+i\eta}\frac{d_{s}^{\mu}Y_{ac\mathbf{k}'}^{s}}{\hbar\omega+\Omega_{s}+i\eta}\right)\\
 & -\frac{\text{e}^{3}}{\hbar^{2}\omega_{1}\omega_{2}}\sum_{s\lambda}\left(-\frac{h_{ba}^{\beta}Y_{cb\mathbf{k}}^{\lambda}d_{\lambda}^{\alpha}}{\hbar\omega_{1}-\Omega_{\lambda}+i\eta}\frac{d_{s}^{\mu*}Y_{ca\mathbf{k}'}^{s*}}{\hbar\omega-\Omega_{s}+i\eta}+\frac{h_{ba}^{\beta}Y_{bc\mathbf{k}}^{\lambda*}d_{\lambda}^{\alpha*}}{\hbar\omega_{1}+\Omega_{\lambda}+i\eta}\frac{d_{s}^{\mu}Y_{ac\mathbf{k}'}^{s}}{\hbar\omega+\Omega_{s}+i\eta}\right)\\
 & -\frac{\text{e}^{3}}{\hbar^{2}\omega_{1}\omega_{2}}\sum_{s\lambda}\left[\frac{h_{ca}^{\mu*}Y_{ba\mathbf{k}}^{s}Y_{bc\mathbf{k}}^{\lambda*}d_{s}^{\alpha}d_{\lambda}^{\beta*}}{(\hbar\omega_{1}-\Omega_{s}+i\eta)(\hbar\omega_{2}+\Omega_{\lambda}+i\eta)}-\frac{h_{ca}^{\mu*}Y_{ab\mathbf{k}}^{s*}Y_{cb\mathbf{k}}^{\lambda}d_{s}^{\alpha*}d_{\lambda}^{\beta}}{(\hbar\omega_{1}+\Omega_{s}+i\eta)(\hbar\omega_{2}-\Omega_{\lambda}+i\eta)}\right]\\
 & -\frac{\text{e}^{3}}{\hbar^{2}\omega_{1}\omega_{2}}\sum_{s\lambda}\left[\frac{h_{ca}^{\mu*}Y_{ba\mathbf{k}}^{s}Y_{bc\mathbf{k}}^{\lambda*}d_{s}^{\beta}d_{\lambda}^{\alpha*}}{(\hbar\omega_{2}-\Omega_{s}+i\eta)(\hbar\omega_{1}+\Omega_{\lambda}+i\eta)}-\frac{h_{ca}^{\mu*}Y_{ab\mathbf{k}}^{s*}Y_{cb\mathbf{k}}^{\lambda}d_{s}^{\beta*}d_{\lambda}^{\alpha}}{(\hbar\omega_{2}+\Omega_{s}+i\eta)(\hbar\omega_{1}-\Omega_{\lambda}+i\eta)}\right].
\end{align*}
Grouping terms with the same denominator, we have
\begin{align*}
 & \sigma_{eh}^{\mu\alpha\beta,4}/C\\
 & =-\sum_{s\lambda}\frac{h_{cb}^{\beta}Y_{ba\mathbf{k}}^{s}d_{s}^{\alpha}}{\hbar\omega_{1}-\Omega_{s}+i\eta}\frac{d_{\lambda}^{\mu*}Y_{ca\mathbf{k}'}^{\lambda*}}{\hbar\omega-\Omega_{\lambda}+i\eta}+\sum_{s\lambda}\frac{h_{ba}^{\beta}Y_{cb\mathbf{k}}^{\lambda}d_{\lambda}^{\alpha}}{\hbar\omega_{1}-\Omega_{\lambda}+i\eta}\frac{d_{s}^{\mu*}Y_{ca\mathbf{k}'}^{s*}}{\hbar\omega-\Omega_{s}+i\eta}\\
 & +\sum_{s\lambda}\frac{h_{cb}^{\beta}Y_{ab\mathbf{k}}^{s*}d_{s}^{\alpha*}}{\hbar\omega_{1}+\Omega_{s}+i\eta}\frac{d_{\lambda}^{\mu}Y_{ac\mathbf{k}'}^{\lambda}}{\hbar\omega+\Omega_{\lambda}+i\eta}-\sum_{s\lambda}\frac{h_{ba}^{\beta}Y_{bc\mathbf{k}}^{\lambda*}d_{\lambda}^{\alpha*}}{\hbar\omega_{1}+\Omega_{\lambda}+i\eta}\frac{d_{s}^{\mu}Y_{ac\mathbf{k}'}^{s}}{\hbar\omega+\Omega_{s}+i\eta}\\
 & -\sum_{s\lambda}\frac{h_{cb}^{\alpha}Y_{ba\mathbf{k}}^{s}d_{s}^{\beta}}{\hbar\omega_{2}-\Omega_{s}+i\eta}\frac{d_{\lambda}^{\mu*}Y_{ca\mathbf{k}'}^{\lambda*}}{\hbar\omega-\Omega_{\lambda}+i\eta}+\sum_{s\lambda}\frac{h_{ba}^{\alpha}Y_{cb\mathbf{k}}^{\lambda}d_{\lambda}^{\beta}}{\hbar\omega_{2}-\Omega_{\lambda}+i\eta}\frac{d_{s}^{\mu*}Y_{ca\mathbf{k}'}^{s*}}{\hbar\omega-\Omega_{s}+i\eta}\\
 & +\sum_{s\lambda}\frac{h_{cb}^{\alpha}Y_{ab\mathbf{k}}^{s*}d_{s}^{\beta*}}{\hbar\omega_{2}+\Omega_{s}+i\eta}\frac{d_{\lambda}^{\mu}Y_{ac\mathbf{k}'}^{\lambda}}{\hbar\omega+\Omega_{\lambda}+i\eta}-\sum_{s\lambda}\frac{h_{ba}^{\alpha}Y_{bc\mathbf{k}}^{\lambda*}d_{\lambda}^{\beta*}}{\hbar\omega_{2}+\Omega_{\lambda}+i\eta}\frac{d_{s}^{\mu}Y_{ac\mathbf{k}'}^{s}}{\hbar\omega+\Omega_{s}+i\eta}\\
 & -\sum_{s\lambda}\frac{h_{ca}^{\mu*}Y_{ba\mathbf{k}}^{s}Y_{bc\mathbf{k}}^{\lambda*}d_{s}^{\alpha}d_{\lambda}^{\beta*}}{(\hbar\omega_{1}-\Omega_{s}+i\eta)(\hbar\omega_{2}+\Omega_{\lambda}+i\eta)}+\sum_{s\lambda}\frac{h_{ca}^{\mu*}Y_{ab\mathbf{k}}^{s*}Y_{cb\mathbf{k}}^{\lambda}d_{s}^{\beta*}d_{\lambda}^{\alpha}}{(\hbar\omega_{2}+\Omega_{s}+i\eta)(\hbar\omega_{1}-\Omega_{\lambda}+i\eta)}\\
 & +\sum_{s\lambda}\frac{h_{ca}^{\mu*}Y_{ab\mathbf{k}}^{s*}Y_{cb\mathbf{k}}^{\lambda}d_{s}^{\alpha*}d_{\lambda}^{\beta}}{(\hbar\omega_{1}+\Omega_{s}+i\eta)(\hbar\omega_{2}-\Omega_{\lambda}+i\eta)}-\sum_{s\lambda}\frac{h_{ca}^{\mu*}Y_{ba\mathbf{k}}^{s}Y_{bc\mathbf{k}}^{\lambda*}d_{s}^{\beta}d_{\lambda}^{\alpha*}}{(\hbar\omega_{2}-\Omega_{s}+i\eta)(\hbar\omega_{1}+\Omega_{\lambda}+i\eta)},
\end{align*}
where $C=\frac{e^{3}}{\hbar^{2}\omega_{1}\omega_{2}}$. 
We note that both $\lambda$ and $s$ are dummy indices so we can redefine them in the summation.
\begin{align*}
 & \sigma_{\text{eh}}^{\mu\alpha\beta,4}/C\\
 & =-\sum_{s\lambda}\frac{d_{s}^{\alpha}d_{\lambda}^{\mu*}(h_{cb}^{\beta}Y_{ba\mathbf{k}}^{s}Y_{ca\mathbf{k}}^{\lambda*}-h_{ba}^{\beta}Y_{cb\mathbf{k}}^{s}Y_{ca\mathbf{k}}^{\lambda*})}{(\hbar\omega_{1}-\Omega_{s}+i\eta)(\hbar\omega-\Omega_{\lambda}+i\eta)}+\sum_{s\lambda}\frac{d_{s}^{\alpha*}d_{\lambda}^{\mu}(h_{cb}^{\beta}Y_{ab\mathbf{k}}^{s*}Y_{ac\mathbf{k}}^{\lambda}-h_{ba}^{\beta}Y_{bc\mathbf{k}}^{s*}Y_{ac\mathbf{k}}^{\lambda})}{(\hbar\omega_{1}+\Omega_{s}+i\eta)(\hbar\omega+\Omega_{\lambda}+i\eta)}\\
 & -\sum_{s\lambda}\frac{d_{s}^{\beta}d_{\lambda}^{\mu*}(h_{cb}^{\alpha}Y_{ba\mathbf{k}}^{s}Y_{ca\mathbf{k}}^{\lambda*}-h_{ba}^{\alpha}Y_{cb\mathbf{k}}^{s}Y_{ca\mathbf{k}}^{\lambda*})}{(\hbar\omega_{2}-\Omega_{s}+i\eta)(\hbar\omega-\Omega_{\lambda}+i\eta)}+\sum_{s\lambda}\frac{d_{s}^{\beta*}d_{\lambda}^{\mu}(h_{cb}^{\alpha}Y_{ab\mathbf{k}}^{s*}Y_{ac\mathbf{k}}^{\lambda}-h_{ba}^{\alpha}Y_{bc\mathbf{k}}^{s*}Y_{ac\mathbf{k}}^{\lambda})}{(\hbar\omega_{2}+\Omega_{s}+i\eta)(\hbar\omega+\Omega_{\lambda}+i\eta)}\\
 & -\sum_{s\lambda}\frac{(h_{ca}^{\mu*}Y_{ba\mathbf{k}}^{s}Y_{bc\mathbf{k}}^{\lambda*}-h_{ca}^{\mu*}Y_{ab\mathbf{k}}^{\lambda*}Y_{cb\mathbf{k}}^{s})d_{s}^{\alpha}d_{\lambda}^{\beta*}}{(\hbar\omega_{1}-\Omega_{s}+i\eta)(\hbar\omega_{2}+\Omega_{\lambda}+i\eta)}+\sum_{s\lambda}\frac{d_{s}^{\alpha*}d_{\lambda}^{\beta}(h_{ca}^{\mu*}Y_{ab\mathbf{k}}^{s*}Y_{cb\mathbf{k}}^{\lambda}-h_{ca}^{\mu*}Y_{ba\mathbf{k}}^{\lambda}Y_{bc\mathbf{k}}^{s*})}{(\hbar\omega_{1}+\Omega_{s}+i\eta)(\hbar\omega_{2}-\Omega_{\lambda}+i\eta)}\\
 & =-\sum_{s\lambda}\frac{d_{s}^{\alpha}d_{\lambda}^{\mu*}\Pi_{\lambda s}^{\beta}}{(\hbar\omega_{1}-\Omega_{s}+i\eta)(\hbar\omega-\Omega_{\lambda}+i\eta)}-\sum_{s\lambda}\frac{d_{s}^{\alpha*}d_{\lambda}^{\mu}\Pi_{s\lambda}^{\beta}}{(\hbar\omega_{1}+\Omega_{s}+i\eta)(\hbar\omega+\Omega_{\lambda}+i\eta)}\\
 & -\sum_{s\lambda}\frac{d_{s}^{\beta}d_{\lambda}^{\mu*}\Pi_{\lambda s}^{\alpha}}{(\hbar\omega_{2}-\Omega_{s}+i\eta)(\hbar\omega-\Omega_{\lambda}+i\eta)}-\sum_{s\lambda}\frac{d_{s}^{\beta*}d_{\lambda}^{\mu}\Pi_{s\lambda}^{\alpha}}{(\hbar\omega_{2}+\Omega_{s}+i\eta)(\hbar\omega+\Omega_{\lambda}+i\eta)}\\
 & +\sum_{s\lambda}\frac{\Pi_{s\lambda}^{\mu*}d_{s}^{\alpha}d_{\lambda}^{\beta*}}{(\hbar\omega_{1}-\Omega_{s}+i\eta)(\hbar\omega_{2}+\Omega_{\lambda}+i\eta)}+\sum_{s\lambda}\frac{d_{s}^{\alpha*}d_{\lambda}^{\beta}\Pi_{\lambda s}^{\mu*}}{(\hbar\omega_{1}+\Omega_{s}+i\eta)(\hbar\omega_{2}-\Omega_{\lambda}+i\eta)},
\end{align*}
where we use Eq.~\ref{eq:inter-exciton coupling} for inter-exciton couplings in the last three lines. It is easy to check that $\Pi^\beta_{\lambda s}=\Pi^{\beta *}_{s\lambda}$.

Finally, combining all three terms and symmetrized $\sigma_{eh}^{\mu\alpha\beta,2}$ and $\sigma_{eh}^{\mu\alpha\beta,3}$, we have
\begin{align*}
 & \sigma_{eh}^{\mu\alpha\beta}(\omega;\omega_{1},\omega_{2})+\sigma_{eh}^{\mu\beta\alpha}(\omega;\omega_{2},\omega_{1})\\
 & =-\frac{\text{e}^{3}}{\hbar^{2}\omega_{1}\omega_{2}}\sum_{\lambda}\left[\left(\frac{d_{\lambda}^{\alpha}d_{\lambda}^{\mu\beta*}}{\hbar\omega_{1}-\Omega_{\lambda}+i\eta}-\frac{d_{\lambda}^{\alpha*}d_{\lambda}^{\mu\beta}}{\hbar\omega_{1}+\Omega_{\lambda}+i\eta}\right)+\left(\frac{d_{\lambda}^{\beta}d_{\lambda}^{\mu\alpha*}}{\hbar\omega_{2}-\Omega_{\lambda}+i\eta}-\frac{d_{\lambda}^{\beta*}d_{\lambda}^{\mu\alpha}}{\hbar\omega_{2}+\Omega_{\lambda}+i\eta}\right)\right]\\
 & -\frac{\text{e}^{3}}{2\hbar^{2}\omega_{1}\omega_{2}}\sum_{\lambda}\left[\left(\frac{d_{\lambda}^{\alpha\beta}d_{\lambda}^{\mu*}}{\hbar\omega-\Omega_{\lambda}+i\eta}-\frac{d_{\lambda}^{\alpha\beta*}d_{\lambda}^{\mu}}{\hbar\omega+\Omega_{\lambda}+i\eta}\right)+\left(\frac{d_{\lambda}^{\beta\alpha}d_{\lambda}^{\mu*}}{\hbar\omega-\Omega_{\lambda}+i\eta}-\frac{d_{\lambda}^{\beta\alpha*}d_{\lambda}^{\mu}}{\hbar\omega+\Omega_{\lambda}+i\eta}\right)\right]\\
 & +\frac{\text{e}^{3}}{\hbar^{2}\omega_{1}\omega_{2}}\sum_{s\lambda}\left[-\frac{d_{\lambda}^{\mu*}\Pi_{\lambda s}^{\beta}d_{s}^{\alpha}}{(\hbar\omega_{1}-\Omega_{s}+i\eta)(\hbar\omega-\Omega_{\lambda}+i\eta)}-\frac{d_{\lambda}^{\mu}\Pi_{s\lambda}^{\beta}d_{s}^{\alpha*}}{(\hbar\omega_{1}+\Omega_{s}+i\eta)(\hbar\omega+\Omega_{\lambda}+i\eta)}\right]\\
 & +\frac{\text{e}^{3}}{\hbar^{2}\omega_{1}\omega_{2}}\sum_{s\lambda}\left[-\frac{d_{\lambda}^{\mu*}\Pi_{\lambda s}^{\alpha}d_{s}^{\beta}}{(\hbar\omega-\Omega_{\lambda}+i\eta)(\hbar\omega_{2}-\Omega_{s}+i\eta)}-\frac{d_{\lambda}^{\mu}\Pi_{s\lambda}^{\alpha}d_{s}^{\beta*}}{(\hbar\omega+\Omega_{\lambda}+i\eta)(\hbar\omega_{2}+\Omega_{s}+i\eta)}\right]\\
 & +\frac{\text{e}^{3}}{\hbar^{2}\omega_{1}\omega_{2}}\sum_{s\lambda}\left[\frac{d_{s}^{\alpha}\Pi_{s\lambda}^{\mu*}d_{\lambda}^{\beta*}}{(\hbar\omega_{1}-\Omega_{s}+i\eta)(\hbar\omega_{2}+\Omega_{\lambda}+i\eta)}+\frac{d_{s}^{\alpha*}\Pi_{s\lambda}^{\mu*}d_{\lambda}^{\beta}}{(\hbar\omega_{1}+\Omega_{s}+i\eta)(\hbar\omega_{2}-\Omega_{\lambda}+i\eta)}\right].
\end{align*}
Without explicit symmetrization, we can write
\begin{align*}
 & \sigma_{eh}^{\mu\alpha\beta}(\omega;\omega_{1},\omega_{2})\\
 & =\frac{-\text{e}^{3}}{\hbar^{2}\omega_{1}\omega_{2}}\sum_{\lambda}\left[\frac{d_{\lambda}^{\alpha}d_{\lambda}^{\mu\beta*}}{\hbar\omega_{1}-\Omega_{\lambda}+i\eta}-\frac{d_{\lambda}^{\alpha*}d_{\lambda}^{\mu\beta}}{\hbar\omega_{1}+\Omega_{\lambda}+i\eta}\right]+\frac{-\text{e}^{3}}{2\hbar^{2}\omega_{1}\omega_{2}}\sum_{\lambda}\left[\frac{d_{\lambda}^{\alpha\beta}d_{\lambda}^{\mu*}}{\hbar\omega-\Omega_{\lambda}+i\eta}-\frac{d_{\lambda}^{\alpha\beta*}d_{\lambda}^{\mu}}{\hbar\omega+\Omega_{\lambda}+i\eta}\right]\\
 & +\frac{\text{e}^{3}}{\hbar^{2}\omega_{1}\omega_{2}}\sum_{s\lambda}\left[-\frac{d_{\lambda}^{\mu*}\Pi_{\lambda s}^{\alpha}d_{s}^{\beta}}{(\hbar\omega-\Omega_{\lambda}+i\eta)(\hbar\omega_{2}-\Omega_{s}+i\eta)}-\frac{d_{\lambda}^{\mu}\Pi_{s\lambda}^{\alpha}d_{s}^{\beta*}}{(\hbar\omega+\Omega_{\lambda}+i\eta)(\hbar\omega_{2}+\Omega_{s}+i\eta)}\right]\\
 & +\frac{\text{e}^{3}}{\hbar^{2}\omega_{1}\omega_{2}}\sum_{s\lambda}\frac{d_{s}^{\alpha}\Pi_{s\lambda}^{\mu*}d_{\lambda}^{\beta*}}{(\hbar\omega_{2}+\Omega_{\lambda}+i\eta)(\hbar\omega_{1}-\Omega_{s}+i\eta)},
\end{align*}
which is the part explicitly shown in Eq.~\ref{eq:2nd_eh} in the main text.

\twocolumngrid

\bibliography{ref}

\end{document}